\documentclass[11pt,a4paper]{article}
\usepackage{jheppub} 

\usepackage{pstricks,pst-coil}

\newcommand{\rme}{{\rm e}}
\newcommand{\rmd}{{\rm d}}

\newcommand{\mre}{\mathrm{e}}
\newcommand{\mrd}{\mathrm{d}}

\newcommand{\bfS}{\mathbf{S}}
\newcommand{\p}{\partial}

\newcommand{\calA}{\mathcal{A}}
\newcommand{\order}[1]{\mathcal{O}\left(#1\right)}

\newcommand{\st}{\mathrm{st}}
\newcommand{\con}{\mathrm{con}}

\newcommand{\m}{\mathrm{m}} 

\newcommand{\fhat}{\hat{f}}

\title{Symanzik effective actions in the large $N$ limit}

\author[a]{J.\ Balog,}
\author[b]{F.\ Niedermayer,}
\author[c]{P.\ Weisz}

\affiliation[a]{Institute for Particle and Nuclear Physics, 
Wigner Research Centre for Physics, \\
MTA Lend\"ulet Holographic QFT Group,
1525 Budapest 114, P.O.B.\ 49, Hungary}
\affiliation[b]{Albert Einstein Center for Fundamental Physics \\
Institute for Theoretical Physics,
Bern University, Sidlerstr.\ 5, 3012 Bern, Switzerland}
\affiliation[c]{Max-Planck-Institut f\"ur Physik, 80805 Munich, Germany}

\emailAdd{balog.janos@wigner.mta.hu}
\emailAdd{niedermayer@itp.unibe.ch}
\emailAdd{pew@mpp.mpg.de}

\abstract{Symanzik effective actions, conjectured to describe 
lattice artifacts, are determined for a class of lattice 
regularizations of the non-linear O($N$) sigma model 
in two dimensions in the leading order of the $1/N$-expansion. 
The class of actions considered includes also ones
which do not have the usual classical limit and
are not (so far) treatable in the framework of ordinary 
perturbation theory. The effective actions obtained
are shown to reproduce previously computed lattice artifacts 
of the step scaling functions defined in finite volume,
giving further confidence in Symanzik's theory of lattice artifacts.
}

\subheader{\hfill MPP-2013-20}

\begin{document} 

\maketitle

\section{Introduction}
\label{Introduction}
Determinations of physical quantities in lattice QCD require
extrapolation of the data to the continuum limit.
For this purpose an ansatz must be made on the nature 
of the lattice artifacts. 
Usually these are assumed to be predominantly 
power-like $\sim a^p$ in the lattice spacing $a$, 
the power $p$ depending on the particular lattice action 
used for the simulation.

The latter ansatz is based on a {\it conjecture} of 
Symanzik \cite{Symanzik} which says that the lattice artifacts 
of correlation functions of the basic fields are  
described by a local effective continuum
Lagrangian involving higher dimensional operators 
which may have only the symmetries of the underlying
lattice theory, and coefficients depending on $a$ 
\footnote{The description of lattice artifacts of 
correlation functions involving composite operators
involve in addition further effective operators.}.

The study of lattice artifacts is not only of theoretical interest:
it is closely related to Symanzik's improvement program 
\cite{Sym83,Sym83a,Lue85,Lue85a}
where the leading cutoff effects are cancelled by adding 
(already on the lattice) $\order{a^2}$
local counterterms with suitably fine-tuned coefficients. Using such
improved actions the lattice artifacts are reduced considerably 
and since the continuum limit is reached faster, 
allows for cheaper MC simulations.

This structural conjecture is based mainly on 
perturbative computations,
both in the coupling and in the $1/N$ expansion, 
in various models.
An all order proof for the existence of Symanzik effective actions
for $\phi^4_4$ theory and $\mathrm{QED}_4$  
has been given by Keller \cite{Keller}, 
albeit this proof is only in the framework of a 
continuum regularization.
However, these theories, at least when regularized on the lattice, 
are probably trivial in the continuum limit.
Renormalized couplings tend to zero as $g\sim c/\ln(a\mu)$ 
and hence the continuum limit (a free theory)
is actually reached only logarithmically!
Treating a renormalized coupling effectively as a constant for a 
range of cutoffs one has for small $g$ a perturbative Lagrangian 
description of the low energy physics, and in this case the Symanzik 
effective Lagrangian describes the leading $a^2$ cutoff corrections to this,
in particular effects of rotation symmetry breaking.

So far an all order perturbative proof of the conjecture has
not been given for asymptotically free theories;
there seems to be no technical obstacle but it is probably
a tedious task. For QCD there is 
support for the validity of the conjecture coming only
from low order perturbative computations, and from the
fact that the numerical data can be convincingly described
by fits of the expected form. 

Asymptotically free theories in 2 dimensions such as the O($N$)
sigma model, offer a laboratory to study cutoff effects 
in more detail, since simulations to large correlation lengths
can be made, and one can also perform studies in the $1/N$-expansion.
We shall restrict attention to this theory in this paper
and the structure of the Symanzik effective action is discussed in
the next subsection. For a class of lattice actions permitting a
standard perturbative formulation it was shown \cite{perturbative} that the 
generic leading cutoff behavior is $\propto a^2(\ln m a)^{N/(N-2)}$
(see subsection~\ref{Artifacts}). 
The additional $\ln^3 a$ factor for $N=3$
becomes visible in precision measurements, and can even mimic 
$\order{a}$ effects in some intermediate range of correlation lengths.

In a recent paper \cite{drastic} a class of models was identified 
with very small lattice artifacts. Some of these models 
are particularly fascinating since they seem to belong to the 
same universality class as the standard-type actions described
above but don't have the usual classical limit and so far a perturbative
treatment for them is unknown (see subsection~\ref{Artifacts}).
However, they can be treated in 
the $1/N$ approximation and there the step scaling functions
(of the finite volume mass gap) were found to be of the form
$a^2\ln^s a$ with $s=2$ in leading order $1/N$ rather than 
$\lim_{N\to\infty} N/(N-2)=1$.
Note that in the framework of Symazik's conjecture, the difference
between the cutoff behavior for different lattice actions
can only arise from the couplings in the effective theory.

In this paper we determine the Symanzik effective action 
for the classes of lattice actions mentioned above in the leading
order of the $1/N$ expansion by matching lattice correlation functions
to the effective theory.
We show how the constructed effective theory reproduces the 
artifacts of the step scaling functions defined in finite volume. 
It is instructive to see how the additional non-perturbative effects 
leading to the $\ln^2 a$ appear in this framework.
The various steps of the computation are given in 
sections~\ref{SymTheo}-\ref{Matching}
and some technical details are relegated to appendices.

\subsection{Symanzik's effective action for $2d$ O$(N)$ sigma models}
\label{EffAct}
As explained above, Symanzik's idea is to mimic the lattice artifacts
(at least up to the leading $\order{a^2}$ order) 
by using an effective Lagrangian in the continuum theory:
\begin{equation}
  -{\cal L}_{{\rm eff}}=-{\cal L}+a^2\sum_i\,c_i(g)\,U_i,
  \label{EffLag}
\end{equation}
where ${\cal L}$ is the original continuum Lagrangian density and in this
sum all operators $U_i$ with engineering dimension $4$ and reflecting the
lattice symmetries have to be taken into account. The c-number coefficients
$c_i(g)$ depend on the lattice action. For perturbative type actions they 
can be calculated in perturbation theory. In this paper we will calculate 
them in the large $N$ expansion.

For practical calculations a useful starting point is the master formula 
of Symanzik's effective theory, which can be obtained using the effective
Lagrangian (\ref{EffLag}) and is written directly in terms of 
correlation functions on the lattice and the corresponding effective 
continuum model:
\begin{equation}
  {\cal G}^X_{{\rm latt}}(\lambda_0,a)=y^r(g)\,{\cal G}^X_{(R)}(g,a^{-1})+
  a^2\,y^r(g)\sum_iv_i(g)\,{\cal G}^X_{i(R)}(g,a^{-1})+\order{a^4} \,.
  \label{master}
\end{equation}
Here ${\cal G}^X_{{\rm latt}}$ represents a general lattice correlation 
function (with $r$ external legs) for any physical quantity $X$, in real space
or momentum space, and ${\cal G}^X_{(R)}$ is the analogous renormalized
quantity calculated in the continuum model. ${\cal G}^X_{{\rm latt}}$ depends
on the set $\lambda_0$ of bare lattice couplings and parameters, while
the continuum correlators depend on the renormalized coupling $g$ and the
renormalization scale $\mu$, which is taken here, for simplicity, 
as the inverse lattice spacing $a^{-1}$. ${\cal G}^X_{i(R)}$ is the
corresponding renormalized continuum correlation function calculated with
insertion of the integral of one of the local dimension 4 operators, $U_i$, 
appearing in Symanzik's effective continuum action. For a given lattice
symmetry the set of operators is fixed and only the c-number coefficient
functions $v_i(g)$, the renormalized coefficients corresponding to
$c_i(g)$, depend on the lattice action. Finally, 
the wave function renormalization constant $y(g)$ and the relation 
$g=g(\lambda_0)$ between the lattice coupling parameters and the continuum 
coupling constant are again action dependent. The latter can be obtained
by calculating some physical mass parameter ${\cal M}$ on both sides:
\begin{equation}
  {\cal M}(g,a^{-1})={\cal M}_{{\rm latt}}(\lambda_0) \,.
\end{equation}

For generating correlation functions we will use the source dependent
action
\begin{equation}
  {\cal A}=\int{\rm d}^Dx\,  {\cal L}
  =\int{\rm d}^Dx
  \left\{\frac{1}{2g_0^2}\,\partial_\mu S\cdot 
    \partial_\mu S-\frac{1}{g_0^2}\,I\cdot S\right\} \,.
\end{equation}
Here $S$ is the $N$-component sigma model field with normalization $S\cdot S=1$
and instead of the bare coupling constant $g_0^2$ we will mainly use the
't Hooft coupling $f_0=Ng_0^2$ (and similarly for the renormalized 
quantities: $f=Ng^2$). In the continuum we will use dimensional 
regularization in $D=2-\varepsilon$ dimensions.

The set of local operators appearing in Symanzik's effective action
always includes the following Lorentz-scalar dimension 4 O$(N)$ invariant
operators:
\begin{equation} \label{O1_O3}
  \begin{split}
    {\cal O}_1&=\frac{1}{8}(\partial_\mu S\cdot\partial_\mu S)^2\,,\\
    {\cal O}_2&=\frac{1}{8}(\partial_\mu S\cdot\partial_\nu S)
    (\partial_\mu S\cdot\partial_\nu S)\,,\\
    {\cal O}_3&=\frac{1}{2}(\Box S\cdot\Box S)\,.
  \end{split}
\end{equation}

These operators mix under renormalization with other Lorentz-scalar operators
and in Symanzik's effective action we also have to include operators which 
are not fully Lorentz scalar but invariant under lattice symmetries only.
The full list of operators in the case of lattices with cubic symmetry
will be given in section~\ref{SymTheo}.

A set of lattice regularized O$(N)$ sigma models with one tunable
coupling $\beta=1/\lambda_0^2$ is given by the following 
quadratic lattice actions:
\begin{equation}
  {\cal A}_{{\rm latt}}=\frac{\beta}{2}\,a^4\sum_{x,y}S(x)\cdot S(y) K(x-y)\,.
  \label{quadratic}
\end{equation}
Here the only requirement is that the Fourier transform of the inverse 
\lq\lq propagator" $K(x)$ defined by
\begin{equation}
  K_p=a^2\sum_x{\rm e}^{-iapx}\,K(x)
\end{equation}
behaves for $a\to0$ as
\begin{equation}
  K_p=p^2+\order{a^2} \,.
\end{equation}
This ensures that the lattice model has the correct classical continuum limit.
For the standard lattice action (ST) we have:
\begin{equation}
  K_p=\hat p^2,\qquad\qquad \hat p_\mu=\frac{2}{a}\sin\frac{ap_\mu}{2}\,.
\end{equation}

An alternative way of writing the master equation is
\begin{equation}
  {\cal G}^X_{{\rm latt}}={\cal G}^{X(0)}(\lambda_0,a)\left\{
    1+a^2\delta^X(\lambda_0,a)+\order{a^4}\right\}\,,
\end{equation}
where the scaling part ${\cal G}^{X(0)}$ is universal (up to wave function
renormalization) and the correction factor is of the form
\begin{equation}
  \delta^X(\lambda_0,a)=\sum_i v_i(g)\,\delta^X_i(g,a)\,.
\end{equation}
Using perturbation theory and renormalization group considerations, it was
shown in \cite{perturbative} that close to the continuum limit
this correction can be represented as 
\begin{equation}
  \delta^X(\lambda_0,a)=C_1\,D_1^X(\Lambda)\left\{
    \tilde\beta^{1+2\chi}+\order{\tilde\beta^{2\chi}} \right\}+
  \order{\tilde\beta^\chi}\,,
\label{resu}
\end{equation}
where $\chi=1/(N-2)$ and close to the continuum
\begin{equation}
  \tilde\beta=\frac{2\pi}{\lambda_0}\sim\ln(\Lambda a)\,.
\end{equation}
In the above formula $C_1$ is a non-universal (action dependent)
constant which is calculable in perturbation theory. On the other hand,
$D_1^X(\Lambda)$ is universal but non-perturbative and depends on the quantity
$X$ in question and on the non-perturbative physical scale~$\Lambda$. 
$D_1^X(\Lambda)$ is related to the non-perturbative matrix element of
the operator with largest anomalous dimension. For $N=3$ (\ref{resu})
predicts a large logarithmic correction proportional to $a^2(\ln a)^3$,
whereas for large $N$ this prediction gives $\order{a^2\ln a}$ corrections.
This is consistent with the known \cite{CaPe,Wolff} $\order{a^2\ln a}$ 
corrections appearing for the standard lattice action 
in the large $N$ limit.

\subsection{Lattice artifacts for non-perturbative actions}
\label{Artifacts}
The universality class of lattice regularized O$(N)$ sigma models is
actually much larger than the set of quadratic actions (\ref{quadratic}).
In \cite{drastic} non-perturbative actions were studied.
For these types of actions the classical continuum limit cannot be expanded 
by usual weak coupling perturbation theory or 
there is no classical limit at all! Nevertheless, as was demonstrated in 
\cite{drastic} the exact quantum continuum limit of these models coincides
with the usual one. Moreover, some of these models show much smaller lattice
artifacts than the conventional ones described by (\ref{quadratic}) type
actions.

The physical quantity studied in \cite{drastic} was the step scaling function
$\sigma(2,u)$ which is defined~\cite{Lue91} as follows. Consider the model 
confined in a periodic (one-dimensional) box of size $L$ and denote by $M(L)$
the mass gap in the box.\footnote{$M$ or $M(L)$ always refers here the 
finite-volume mass gap, while we denote the infinite-volume
mass gap by $m$.}
Introducing the dimensionless LWW coupling \cite{Lue91}
\begin{equation}
  u=L\,M(L),
\end{equation}
the step scaling function
\begin{equation}
  \sigma(2,u)=2L \, M(2L)
\end{equation}
describes how the LWW coupling changes under doubling the size of the box 
(a discrete renormalization group transformation). On the lattice we have 
\begin{equation}
  u=L \,M(L,a),\qquad\qquad \Sigma(2,u,a/L)= 2L \,M(2L,a).
\end{equation}
We will also use the notations
\begin{equation}
  u^\prime=\Sigma(2,u,a/L),\qquad\qquad 
  u_\infty^\prime=\sigma(2,u)=\Sigma(2,u,0)\,.
\end{equation}

For large $N$ the step scaling function is known exactly \cite{Lue82c}.
It is given by the solution of the implicit equation
\begin{equation} \label{f0u2}
  f_0(u)=f_0(u_\infty^\prime)+\frac{\ln2}{2\pi}\,.
\end{equation}
The function $f_0(u)$ will be given explicitly by \eqref{f0u_def}. 

Let us introduce the variable\footnote{In \cite{drastic} 
it was defined as $z=f_0(u) + \ln(L/a)/(2\pi)$.
Using eqs.~\eqref{f0us} and \eqref{f0u_as} one can show
that the two definitions coincide. Here we want to stress that 
$z$ depends only on the lattice spacing and not on the physical
volume.} 
\begin{equation} \label{zdef}
  z = \frac{1}{2\pi} \ln \left( \frac{\sqrt{32}}{ma} \right) \,.
\end{equation}
($\xi = 1/(ma)$ is the correlation length in an infinite volume
measured in lattice units.)

The $\order{a^2}$ lattice artifacts for the step scaling function 
can be written as
\begin{equation} \label{uprime}
  u^\prime=u_\infty^\prime+\frac{a^2}{L^2}\nu(u,z)+\order{a^4}\,,
\end{equation}
where the dependence of the correction coefficient $\nu(u,z)$ on $a/L$
must be weak (logarithmic).

It was found in \cite{drastic} that in terms of this variable the coefficient
function $\nu(u,z)$ for the standard action is of the
form\footnote{The functions $t_i(u)$ together with the similarly defined
functions $\bar t_i(u)$ and $T_i(u)$ 
will be given in section~\ref{LattAct}.
}
\begin{equation}
  \nu_{\rm ST}(u,z)=t_0(u)+t_1(u)\,z \,. 
\end{equation}
This is consistent with the $\ln a$ behavior discussed above.
For the non-perturbative constrained (con) and mixed (mix) lattice actions 
described in section~\ref{LattAct}
the corresponding coefficient functions are given by
\begin{equation}
  \nu_{{\rm con}}(u,z)=\bar t_0(u)+\bar t_1(u)\,z+\bar t_2(u)\,z^2 \,,
\end{equation}
and
\begin{equation} \label{nu_mix}
  \nu_{{\rm mix}}(u,z)=T_0(u)+T_1(u)\,z+T_2(u)\,z^2 \,,
\end{equation}
respectively.

The $(\ln a)^2$ behavior found for the non-perturbative actions is in apparent
contradiction with the result (\ref{resu}) found for quadratic (perturbative) 
actions. But one can notice (see section~\ref{LattAct}) the structure
\begin{equation}
  \begin{split}
    t_1(u)&=-\frac{1}{8f_0^\prime(u_\infty^\prime)}\left(u^2-
      \frac{1}{4}u_\infty^{\prime2}\right)\,,\\
    \bar t_2(u)&=-8t_1(u)\,,\\
    T_2(u)&\propto t_1(u)\,.
  \end{split}
\end{equation}
This structure hints at
the fact that here $t_1(u)$, $\bar t_2(u)$ and $T_2(u)$ are all proportional
to the matrix element of the same operator and only the behavior of the
coefficient function (proportional to $z$ for ST and to $z^2$ for con and mix)
is different.

The purpose of this paper is to show by explicitly constructing the large $N$
effective action and calculating the relevant matrix elements that the
structure described above does indeed hold and more generally that Symanzik's
effective action description is valid also beyond perturbation theory.

\subsection{Symanzik's strategy}
\label{strategy}
To verify Symanzik's effective action approach for large $N$ O$(N)$ sigma 
models using the step scaling function we have to go through the following 
steps: 
\begin{enumerate}
\item
  Calculate the coefficients in the master formula (\ref{master}) by calculating
  the 2-point and 4-point correlation functions in infinite volume on both
  sides and comparing them. In the continuum we need the ordinary correlation 
  functions and also correlation functions with dimension 4 operators appearing 
  in the effective action inserted. On the lattice side we need the correlation 
  functions expanded near the continuum limit up to $\order{a^2}$ precision.
\item
  Calculate the 2-point function (with and without operator insertion) in the
  continuum in finite volume. From this one can calculate the corrections to
  the mass gap and the step scaling function.
\item
  Compare the step scaling function directly calculated on the lattice
  with the ones obtained from Symanzik's effective action in step 2 using
  the coefficients calculated in step 1 and verify matching.
\end{enumerate}

The dimensional regularization calculations necessary for steps 1 and 2
will be presented in section~\ref{SymTheo}. The lattice calculations and the
matching of the results with Symanzik's effective theory will be presented
in section~\ref{LattAct} and section~\ref{Matching}, respectively.

\section{Symanzik theory for the large $N$ sigma model}
\label{SymTheo}
In this section we give the results for those continuum correlation functions 
of the O$(N)$ sigma model which are necessary to construct Symanzik's 
effective action in the large $N$ limit. We will use dimensional 
regularization in $2-\varepsilon$ dimensions.

We start by recalling the Feynman rules for the large $N$ expansion for sigma
models in Fourier space:
\begin{align}
  \intertext{$S^aS^b$ (sigma field) propagator:}
  &
  \psset{unit=0.5mm}
  \begin{pspicture}(0,60)(90,80)
    \psline[linewidth=1pt]{-}(4,70)(48,70)
    \rput(1,70){$a$}
    \rput(51,70){$b$}
    \rput(80,70){$\displaystyle \frac{\delta^{ab}}{p^2+m^2}$}
  \end{pspicture}
  \\
  \intertext{$\lambda\lambda$ (auxiliary field) propagator:}
  &
  \psset{unit=0.5mm}
  \begin{pspicture}(0,40)(90,60)
    \rput(4,50){%
      \psCoil[coilaspect=0,coilheight=1.5,coilwidth=3]{90}{3690}
    }
    \rput(80,50){$\displaystyle \frac{-2}{N B(p)}$}
  \end{pspicture}
  \\
  \intertext{$S^aS^b\lambda$ vertex:}
  &
  \psset{unit=0.5mm}
  \begin{pspicture}(0,6)(90,35)
    \rput(4,20){%
      \psCoil[coilaspect=0,coilheight=1.5,coilwidth=3]{90}{1890}
    }
    \psline{-}(27.6,20)(50,30)
    \psline{-}(27.6,20)(50,10)
    \rput(53,30){$a$}
    \rput(53,10){$b$}
    \rput(80,20){$\delta^{ab}$}
  \end{pspicture}
\end{align}
Here the function $B(p)$ is given by the simple 1-loop integral
\begin{equation}
  B(p)=\int\frac{{\rm d}^Dk}{(2\pi)^D}\,\frac{1}{(k^2+m^2)[m^2+(k+p)^2]}.
  \label{Bp}
\end{equation}
Finally, correlation functions with $r$ external sigma lines constructed by the
above rules have to be multiplied by the overall factor 
\begin{equation}
  \left(\frac{f_0}{N}\right)^{r/2} \,,
\end{equation}
and there is an extra $1/2$ factor for every closed
sigma loop.\footnote{Note that loops formed from the legs of our
local operators do not count in this rule. All our diagrams
shown in figures~\ref{F1}-\ref{F10} are \lq\lq tree" diagrams from this point
of view and only the one corresponding to figure~\ref{F11} is a one-loop 
diagram. The $1/2$ rule applies for this last case only.}
Only sigma loops with at least three $SS\lambda$
vertices need to be considered.

Next we give the complete list of local operators necessary to construct 
Symanzik's effective action for lattices with cubic symmetry. In addition to
the Lorentz scalar operators ${\cal O}_1$, ${\cal O}_2$, ${\cal O}_3$ already
defined in subsection~\ref{EffAct}, we need the following ones.
\begin{align}
  U_3 & = -\frac{N}{4f_0}\,I\cdot\Box S,\\
  U_4 & = \frac{N}{8f_0}(I\cdot S)^2-\frac{I\cdot I}{8f_0},\\
  U_5 & = \frac{N}{2f_0}\,I\cdot I,\\
  U_6 & = \frac{N}{f_0}\,\left\{-\frac{6}{D+2}{\cal O}_3+\sum_{\mu=1}^D
    \,S\cdot\partial_\mu^4S\right\},\label{U6def}\\
  U_7 & = \frac{N}{f_0}\left\{-\frac{8}{D+2}\left({\cal O}_1+2{\cal O}_2\right)
    +\sum_{\mu=1}^D\,\left(\partial_\mu S\cdot\partial_\mu S\right)^2\right\},
\label{U7def}\\
  U_8 & = \frac{N}{4f_0}I\cdot S,\\
  U_9 & = \frac{N}{2f_0}(\partial_\mu S\cdot\partial_\mu S)-2U_8.\label{U9def}
\end{align}
We will use as our operator basis the operators $U_3,\dots,U_9$ just defined
above and the combinations
\begin{equation}
  U_1=\frac{N}{f_0}{\cal O}_1+\frac{1}{2}U_5-U_3-U_4,\qquad\qquad
  U_2=\frac{N}{f_0}\left({\cal O}_1-D{\cal O}_2\right).\label{U12def}
\end{equation}
$U_3$, $U_4$ and $U_5$ are Lorentz-scalar but source-dependent operators,
while $U_6$ and $U_7$ are components of traceless Lorentz-tensor operators 
invariant under the discrete (hyper-)cubic rotation symmetry group only. 
As explained in 
\cite{perturbative}, it is necessary to include them in the set of operators 
in the effective action. Here the set is enlarged by the dimension 2 operators
$U_8$ and $U_9$. Since their matrix elements are proportional to the 
dynamically generated mass $m^2$, they were not present in the perturbative
treatment of \cite{perturbative}, but need to be included here.

\subsubsection*{2-point functions}

We will use the following notation for connected 2-point functions:
\begin{align}
  \langle S^a(x)S^a(y)\rangle_c & = {\cal G}^{(2)}(x,y) \,, \\
  \langle K_i S^a(x)S^a(y)\rangle_c & = {\cal G}^{(2)}_i(x,y) \,.
\end{align}
Here $K_i$ stands for the integrated (zero momentum) operator
\begin{equation}
  K_i=\int{\rm d}^Dz\,U_i(z).
\end{equation}
In Fourier space we define\footnote{also similarly without the 
index $i$}
\begin{equation}
  \int {\rm d}^Dx\,\int {\rm d}^Dy\,{\rm e}^{ixq_1}\,{\rm e}^{iyq_2}\,
  {\cal G}^{(2)}_i(x,y)=(2\pi)^D\,\delta(q_1+q_2)\,\tilde{\cal G}^{(2)}_i(k),
\end{equation}
where $q_1=-q_2=k$ and we will also use the \lq\lq amputated'' 2-point 
functions $G^{(2)}_i(k)$ defined by
\begin{equation}
  \tilde{\cal G}^{(2)}_i(k)=\frac{G^{(2)}_i(k)}{(k^2+m^2)^2}.
\end{equation}
An important consequence of the normalization condition $S\cdot S=1$ is that 
\begin{equation}
  {\cal G}^{(2)}(x,x)=1
\end{equation}
and this can be rewritten in Fourier space as the gap equation
to leading order in $1/N$
\begin{equation} \label{gap}
  \frac{1}{f_0}=\int\frac{{\rm d}^Dk}{(2\pi)^D}\,\frac{1}{k^2+m^2}.
\end{equation}

\subsubsection*{4-point functions}

For connected 4-point functions we will use the analogous notations:
\begin{align}
  \langle S^a(x)S^a(y)S^b(w)S^b(z)\rangle_c & ={\cal G}^{(4)}(x,y,w,z) \,, \\
  \langle K_i S^a(x)S^a(y)S^b(w)S^b(z)\rangle_c & ={\cal G}^{(4)}_i(x,y,w,z) \,,
\end{align}
\begin{equation}
  \begin{split}
    \int {\rm d}^Dx\,\int {\rm d}^Dy\,
    \int {\rm d}^Dw\,\int {\rm d}^D&z\,
    {\rm e}^{ixp_1}\,{\rm e}^{iyp_2}\,
    {\rm e}^{iwq_1}\,{\rm e}^{izq_2}\,
    {\cal G}^{(4)}_i(x,y,w,z)\\
    &=(2\pi)^D\,\delta(p_1+p_2+q_1+q_2)\,
    \tilde{\cal G}^{(4)}_i(p_1,p_2,q_1,q_2),
  \end{split}
\end{equation}
\begin{equation}
  \tilde{\cal G}^{(4)}_i(p_1,p_2,q_1,q_2)=\frac{G^{(4)}_i(p_1,p_2,q_1,q_2)}
  {(p_1^2+m^2) (p_2^2+m^2) (q_1^2+m^2) (q_2^2+m^2)}.
\end{equation}

\subsubsection*{Renormalization}

We will have to perform the usual coupling constant and wave function 
renormalization, which in our problem has to be supplemented by the 
renormalization of local operators. We introduce the notation $f$
for the renormalized 't Hooft coupling and $G^{(r)}_{i(R)}$, 
$G^{(r)}_{(R)}$ for renormalized (amputated) $r$-point functions with and 
without operator insertion. For the coupling renormalization we can use
the gap equation (\ref{gap}) and we find 
\begin{equation}
  f_0=\mu^\varepsilon Z(f,\varepsilon) f,
  \label{couplingren}
\end{equation}
where
\begin{equation}
  \frac{1}{Z}=1+\frac{f}{2\pi\varepsilon}+\order{1/N}
  \label{wfZ}
\end{equation}
and
\begin{equation}
  \frac{1}{f}=\frac{\gamma}{4\pi}+\frac{1}{2\pi}\ln\frac{\mu}{m}.
  \label{renormf}
\end{equation}
Here and below the constant $\gamma$ is given by
\begin{equation}
  \gamma=\ln4\pi+\Gamma^\prime(1).
\end{equation}
The renormalized coupling depends on the renormalization scale $\mu$, which
we will eventually identify, for simplicity, with the inverse lattice spacing 
of the lattice model. 

It turns out that the wave function renormalization constant $Z$ is 
to leading order identical
with the one appearing in (\ref{couplingren}) and (\ref{wfZ}) and all our 
operators $U_i$ renormalize diagonally in the large $N$ limit. If we define
the operator renormalization constants $Z_i$ by
\begin{equation}
  G^{(r)}_{(R)}=Z^{-r/2}G^{(r)},\qquad\qquad
  G^{(r)}_{i(R)}=Z^{-r/2}Z_iG^{(r)}_i
\end{equation}
we have
\begin{equation}
  Z_1=Z_9=Z,\qquad\quad Z_2=Z_4=Z_7=\frac{1}{Z},\qquad\quad Z_3=Z_5=Z_6=Z_8=1.
\end{equation}

\subsubsection*{Renormalized 2-point functions}

For the construction of Symanik's effective action we will need the
dimensionally regularized, renormalized 2-point functions with and without
insertion of the (integrated) local operators $U_i$. The details of the
calculation will be presented in appendix A. Here we just give a list of the 
results for the infinite volume renormalized 2-point functions.
\begin{align}
  G^{(2)}_{(R)}  & = f(k^2+m^2)\,, \label{G2} \\
  G^{(2)}_{1(R)} & = 2\pi m^4 \,, \label{G2_1} \\
  G^{(2)}_{2(R)} & = 0 \,, \label{G2_2} \\
  G^{(2)}_{3(R)} & = \frac{f}{2}k^2(k^2+m^2) \,, \label{G2_3} \\
  G^{(2)}_{4(R)} & = 0 \,, \label{G2_4} \\
  G^{(2)}_{5(R)} & = f(k^2+m^2)^2 \,, \label{G2_5} \\
  G^{(2)}_{6(R)} & = 2f\left(k^4-\frac34 (k^2)^2\right) \,, \label{G2_6} \\
  G^{(2)}_{7(R)} & = 0 \,, \label{G2_7} \\
  G^{(2)}_{8(R)} & = \frac{f}{2}(k^2+m^2) \,, \label{G2_8} \\
  G^{(2)}_{9(R)} & = -4\pi m^2. \label{G2_9} 
\end{align}
Here the notation
\begin{equation}
k^4=\sum_{\mu=1,2}k_\mu^4
\end{equation}
is introduced for two-dimensional vectors.

\subsubsection*{Renormalized 4-point functions}

The calculation of the renormalized 4-point functions will be presented in
appendix B. Here we just list the results of the calculation.
Below $p=p_1+p_2=-(q_1+q_2)$.
\begin{align}
  G^{(4)}_{(R)} & = \frac{-2f^2}{Nb(p)} \,, \label{G4} \\
  G^{(4)}_{1(R)} & = \frac{f}{Nb^2(p)}-\frac{m^2f}{N}X \,, \label{G4_1} \\
  G^{(4)}_{2(R)} & = \frac{f^3}{N}\Big\{-Q^2+Q\left(p_1p_2+q_1q_2
    -\frac{2(p_1p)(p_2p)}{p^2}-\frac{2(q_1p)(q_2p)}{p^2}\right)\nonumber\\
  & \qquad\qquad +(p_1p_2)(q_1q_2)-(p_1q_1)(p_2q_2)-(p_1q_2)(p_2q_1)\Big\}
  \,,\label{G4_2} \\
  G^{(4)}_{3(R)} & = -\frac{f^2}{2Nb(p)}(p_1^2+p_2^2+q_1^2+q_2^2) \,,
  \label{G4_3} \\
  G^{(4)}_{4(R)} & = \frac{f^3}{N}\left(\frac{p_1^2+p_2^2}{2}+m^2\right)
  \left(\frac{q_1^2+q_2^2}{2}+m^2\right) \,,\label{G4_4} \\
  G^{(4)}_{5(R)} & = 0 \,,\label{G4_5} \\
  G^{(4)}_{6(R)} & = -\frac{4f^2}{Nb(p)}
  \left\{\frac{p_{1}^4}{p_1^2+m^2}+\frac{p_{2}^4}{p_2^2+m^2}+
    \frac{q_{1}^4}{q_1^2+m^2}
    +\frac{q_{2}^4}{q_2^2+m^2}\right\}\nonumber \\
  & +\frac{3f^2}{Nb(p)}(p_1^2+p_2^2+q_1^2+q_2^2)+
  \frac{3m^2f^2}{4\pi N}X+\frac{8f^2}{Nb^2(p)}Y \,,\label{G4_6} \\
  G^{(4)}_{7(R)} & = \frac{8f^3}{N}\Big\{Q^2\Big(
-\frac{3}{4}+\frac{p^4}{(p^2)^2}
    \Big)+Q\Big(-\frac{p_1p_2+q_1q_2}{4}-\frac{(p_1p)(p_2p)+
      (q_1p)(q_2p)}{2p^2}
\nonumber\\
  & \qquad\qquad +\sum_\mu\frac{p_\mu^2(p_{1\mu}p_{2\mu}+q_{1\mu}
    q_{2\mu})}{p^2}\Big)+\sum_\mu p_{1\mu}p_{2\mu}q_{1\mu}q_{2\mu}\nonumber\\
  & \qquad\qquad -\frac{(p_1p_2)(q_1q_2)+(p_1q_1)(p_2q_2)+
    (p_1q_2)(p_2q_1)}{4}\Big\} \,,\label{G4_7} \\
  G^{(4)}_{8(R)} & = -\frac{2f^2}{Nb(p)} \,,\label{G4_8} \\
  G^{(4)}_{9(R)} & = \frac{2f}{N}X \label{G4_9}  .
\end{align}
Here we introduced the 2-dimensional limit of the (finite) loop integral
(\ref{Bp}):
\begin{equation}
  b(p)=\int\frac{{\rm d}^2k}{(2\pi)^2}\,\frac{1}{(k^2+m^2)[m^2+(k+p)^2]},
  \label{bp}
\end{equation}
and used the notation
\begin{equation} \label{Qdef}
  Q=m^2-\frac{1}{4\pi b(p)} \,,
\end{equation}
\begin{equation}
  X=\frac{4\pi m^2}{b(p)}\left\{\frac{1}{b(p)}\frac{\partial b(p)}{\partial m^2}
    +\frac{1}{p_1^2+m^2}+\frac{1}{p_2^2+m^2}+\frac{1}{q_1^2+m^2}+
    \frac{1}{q_2^2+m^2}\right\} \,,
\label{renX}
\end{equation}
\begin{equation}
  \begin{split}
    Y&=-3m^2b(p)+\frac{3p^2}{4}
    \left(m^2+\frac{p^2}{4}\right)\frac{\partial b(p)}
    {\partial m^2}+\frac{3}{32\pi}\left(5+\frac{p^2}{m^2}\right)\\
    &+\frac{p^4}{(p^2)^2}\left\{
      2m^2b(p)-\left(\frac{m^4}{2}+m^2p^2+\frac{(p^2)^2}{4}\right)
      \frac{\partial b(p)}{\partial m^2}-\frac{1}{8\pi}
      \left(5+\frac{p^2}{m^2}\right) \right\}.
  \end{split}
\label{renY}
\end{equation}

\subsubsection*{2-point function in strip geometry}

After we determine the coefficients in Symanzik's effective action we can
calculate the lattice correction to the mass gap and the step scaling function
from the effective action if we know the 2-point functions (with and without
operator insertions) in finite volume. In this subsection $M$ is denoting the
finite volume mass $M(L)$. (The mass parameter used in the infinite
volume considerations is $m=M(\infty)$). In two dimensions space is finite 
(periodic with period $L$) in the 1 direction and infinite in the 2 (time)
direction corresponding to a discrete spectrum of the momentum component $k_1$
and continuous spectrum for the momentum component $k_2$. In $D$ dimensions
we will assume that the spectrum of $k_1$ is discrete and all other components
$k_2,\dots,k_D$ are continuous. We checked that the $D\to2$ results remain 
the same if we assume that in $D$ dimensions only the time component $k_2$ is 
continuous and all space components $k_\mu$ ($\mu=1,3,\dots,D$) are discrete.
Finite volume quantities will be indicated by an overline. The details of 
the calculation can be found in appendix A. Here we only present the results.

\begin{align}
  \overline{G}^{(2)}_{(R)} & = f(k^2+M^2) \,,\label{fin20}\\
  \overline{G}^{(2)}_{1(R)} & = \frac{M^2}{2\overline{b}(0)} \,,\\
  \overline{G}^{(2)}_{2(R)} & = \frac{f^2h}{2}\left(k_2^2-k_1^2+\frac{\ell}
    {\overline{b}(0)}\right) \,,\\
  \overline{G}^{(2)}_{3(R)} & = \frac{f}{2}k^2(k^2+M^2) \,,\\
  \overline{G}^{(2)}_{4(R)} & = 0 \,,\\
  \overline{G}^{(2)}_{5(R)} & = f(k^2+M^2)^2 \,,\\
  \overline{G}^{(2)}_{6(R)} & = 2fk^4
  +\frac{3f}{2}\left(M^4-(k^2)^2\right)
  -\frac{2f\bar{\rho}}{\overline{b}(0)} \,,\\
  \overline{G}^{(7)}_{2(R)} & = f^2h\left(k_1^2-k_2^2-\frac{\ell}
    {\overline{b}(0)}\right) \,,\\
  \overline{G}^{(2)}_{8(R)} & = \frac{f}{2}(k^2+M^2) \,,\\
  \overline{G}^{(2)}_{9(R)} & = -\frac{1}{\overline{b}(0)}.\label{fin29}
\end{align}
Here we introduced the finite volume integrals
\begin{equation}
  h=\int_{(L)}\frac{{\rm d}^2q}{(2\pi)^2}\,\frac{q_1^2-q_2^2}{q^2+M^2}\qquad
  \qquad{\rm and} \qquad\qquad
\ell=\int_{(L)}\frac{{\rm d}^2q}{(2\pi)^2}\,\frac{q_1^2-q_2^2}{(q^2+M^2)^2} \,,
\label{hell}
\end{equation}
where the meaning of the integration symbol
\begin{equation}
  \int_{(L)}\frac{{\rm d}^2q}{(2\pi)^2}
\end{equation}
is an integration over the variable $q_2$ and sum over the quantized $q_1$.
The above two integrals are finite (after regularization). The parameter
$\bar \rho$ is defined by a regularized (divergent) finite volume integral
\begin{equation}
  \sum_{\mu=1}^D\int_{(L)}\frac{{\rm d}^Dq}{(2\pi)^D}\,
  \frac{q_\mu^4}{(q^2+M^2)^2}
  +\frac{6M^2}{D+2}\,\frac{1}{f_0}=\bar \rho+\order{\varepsilon}.
\label{barrho}
\end{equation}

\subsubsection*{The finite volume gap equation and the step scaling function}

The gap equation, which follows from the normalization $S\cdot S=1$, is of
the same form in finite volume as (\ref{gap}):
\begin{equation}
  \frac{1}{f_0}=\int\frac{{\rm d}^Dk}{(2\pi)^D}\,\frac{1}{k^2+m^2}=
  \int_{(L)}\frac{{\rm d}^Dk}{(2\pi)^D}\,\frac{1}{k^2+M^2}.
  \label{gapM}
\end{equation}
From this one can calculate \cite{Lue82c} the (finite) relation between the 
infinite volume and finite volume mass parameters:
\begin{equation}
  \ln\frac{m}{M}+F(u)=0 \,,
\label{fvm}
\end{equation}
where $u=LM$ and
\begin{equation}
  F(u)=\int_0^\infty\frac{{\rm d}t}{t}\,{\rm e}^{-tu^2}\left(\sum_{k=1}^\infty
    {\rm e}^{-\frac{k^2}{4t}}\right).
\label{Fdef}
\end{equation}
An alternative representation of the function $F(u)$ is given by
\cite{CaPe}
\begin{equation}
  F(u)=\frac{\pi}{u}+\ln u -\gamma
  +G_0\left(\frac{u}{2\pi}\right) \,,
\end{equation}
where
\begin{equation}
  G_0(w)=\sum_{k=1}^\infty\left\{\frac{1}{\sqrt{k^2+w^2}}-\frac{1}{k}\right\}.
\end{equation}
We will often use the function\footnote{The function $f_0(u)$ must not be 
confused with the bare coupling $f_0$, which has no argument.}
$f_0(u)$ defined by
\begin{equation} \label{f0u_def}
  f_0(u)=\frac{1}{2\pi}\left\{F(u)-\ln u+\ln\sqrt{32}\right\} \,,
\end{equation}
in terms of which the step scaling transformation for a general
scale $s$ can be written as
\begin{equation}  \label{f0us}
  f_0(u)=f_0(\sigma(s,u))+\frac{\ln s}{2\pi}.
\end{equation}

For further reference note that $f_0(u)$ has the asymptotic 
form \cite{drastic}
\begin{equation} \label{f0u_as}
  f_0(u) = \frac{1}{2\pi} \left\{ -\ln u + \ln \sqrt{32}\right\} + 
  \order{\mre^{-u}} \,.
\end{equation}

The finite volume version of the integral (\ref{bp}) at zero momentum
can also be expressed in terms of $f_0$:
\begin{equation}
  \overline{b}(0)=-\frac{uf_0^\prime(u)}{2M^2}.
\end{equation}
We will need a second function defined similarly to $F(u)$:
\begin{equation}
  H(u)=-\frac{1}{2}\int_0^\infty\frac{{\rm d}t}{t^2}\,
  {\rm e}^{-tu^2}\left(\sum_{k=1}^\infty {\rm e}^{-\frac{k^2}{4t}}\right).
\label{Hdef}
\end{equation}
This function has the alternative representation
\begin{equation}
  H(u)=\pi u+\frac{u^2}{2}\ln u
  -\frac14 (1 + 2\gamma) u^2
  +4\pi^2 G_1\left(\frac{u}{2\pi}\right)-\frac{\pi^2}{3} \,,
\end{equation}
where
\begin{equation}
  G_1(w)=\sum_{k=1}^\infty\left\{\sqrt{k^2+w^2}-k-\frac{w^2}{2k}\right\}.
\end{equation}

The rest of the finite volume parameters appearing in the finite volume
2-point functions \eqref{fin20}-\eqref{fin29} can be given in terms of these
functions:
\begin{equation}
  \ell=\frac{uF^\prime(u)}{4\pi}=\frac{1}{4\pi}+\frac{u}{2}f_0^\prime(u) \,,
  \qquad\qquad
  h=-\frac{M^2}{2\pi}\,F(u)+\frac{M^2H(u)}{\pi u^2}
\label{landh}
\end{equation}
and
\begin{equation}
  \bar\rho=\frac{3M^2}{16\pi}-M^2\ell-\frac{h}{2}.
\label{barrhores}
\end{equation}

For the cutoff dependence we will also need the functions $f_1(u)$
and $f_2(u)$ \cite{drastic}:
\begin{align}
  f_1(u) & = \frac{\pi}{6}\left[ \frac{1}{12} 
    - G_1\left( \frac{u}{2\pi}\right) \right]
  -\frac{u}{12} - \frac{u^2}{16\pi}
  \left[ k + \frac23 G_0\left( \frac{u}{2\pi}\right) -\frac13 \right] \,,
  \label{f1u} \\
  f_2(u) & = 2\pi G_1\left( \frac{u}{2\pi}\right) -\frac{\pi}{6}
  +\frac{u}{2} + \frac{u^2}{8\pi} (2k-1) \,,
  \label{f2u} 
\end{align}
where $k = -\Gamma'(1) -\ln\pi + \frac12 \ln 2$.

\subsubsection*{Master equation for 2-point function in the strip geometry}

Let us now apply Symanzik's method to calculate the lattice step scaling 
function by writing the master equation (\ref{master}) for the 2-point function
in finite volume:
\begin{equation}
  \tilde{\cal G}^{(2)}_{{\rm latt}}=y^2\left\{\overline{{\cal G}}^{(2)}_{(R)}+
    a^2\sum_{i=1}^9v_i\,\overline{{\cal G}}^{(2)}_{i(R)}+\order{a^4}\right\}.
\end{equation}
From here the lattice correction to the finite volume mass gap is calculated as
\begin{equation}
  M^2_{{\rm latt}}=M^2-\frac{a^2}{f}\sum_{i=1}^9\,v_i\,
  \label{Mlatt}
  \overline{g}_i+\order{a^4} \,,
\end{equation}
where the coefficients $\overline{g}_i$ are 
\begin{equation} \label{gbar_i}
  \overline{g}_i=\overline{G}^{(2)}_{i(R)}(0,iM).
\end{equation}
If we take $\mu=1/a$, i.e. identify the renormalization scale $\mu$ with the 
inverse lattice spacing, we can give the connection between the renormalized
't Hooft coupling $f$ and the lattice parameters using (\ref{renormf}):
\begin{equation} \label{ln_xi_DR}
  \ln\xi=\frac{2\pi}{f}-\frac{\gamma}{2} \,,
\end{equation}
where $\xi$ is the lattice correlation length in infinite volume.

Using the results of the previous subsection, the coefficients in (\ref{Mlatt})
are
\begin{align}
  \overline{g}_1 & = \frac{M^2}{2\overline{b}(0)} \,, \label{gbar_1} \\
  \overline{g}_2 & = \frac{f^2h}{2}\left\{\frac{\ell}{\overline{b}(0)}-
    M^2\right\} \,, \label{gbar_2} \\
  \overline{g}_3 & = \overline{g}_4=\overline{g}_5=\overline{g}_8=0 \,,
  \label{gbar_3458}\\
  \overline{g}_6 & = 2fM^4-\frac{2f\bar\rho}{\overline{b}(0)} \,, 
  \label{gbar_6}\\
  \overline{g}_7 & = -2\,\overline{g}_2 \,, \label{gbar_7} \\
  \overline{g}_9 & = -\frac{1}{\overline{b}(0)} \,.  \label{gbar_9}
\end{align}

\section{Lattice actions}
\label{LattAct}
We consider here the large $N$ limit of the standard, the constrained and the
mixed lattice actions. 
Since the standard and the constrained actions can be obtained
as special cases of the mixed action, we shall discuss in detail only the
latter one.
\subsection{Standard action}
The action is given by
\begin{equation} \label{Ast} \calA_\st[\bfS] = \frac{\beta_\st}{2}
  \sum_{x,\mu} (\p_\mu \bfS_x)^2 \,,
\end{equation}
where $\bfS_x^2=1$ and $\p_\mu$ denotes the forward lattice derivative.

As usual, the large $N$ limit is taken as
\begin{equation}
  \beta_\st = \frac{N}{f_\st}\,,\qquad f_\st=\mathrm{fixed} \,.
\end{equation}
Note for further reference that for large $N$ the distribution of 
$(\p_\mu \bfS_x)^2$ approaches a $\delta$-function at 
$(\p_\mu \bfS_x)^2=f_\st/2$ (for $D=2$ Euclidean dimensions), as can be 
shown in perturbation theory.

For the standard action the infinite volume mass gap $m$ is given by 
\cite{Lue82c}
\begin{equation} \label{m_st} 
  \frac{1}{f_\st} = z \Big[ 1 + \order{m^2a^2} \Big] \,,
\end{equation}
where $z$ is given by eq.~\eqref{zdef}.

\subsection{Constrained action}
The action is the same as the standard one except that the configurations are
restricted by the constraint $(\p_\mu \bfS_x)^2 < \epsilon$:
\begin{equation} \label{Acon} \calA_\con[\bfS] =
  \begin{cases}
    \frac12 \beta_\con \sum_{x,\mu} (\p_\mu \bfS_x)^2 & \text{for $(\p_\mu
      \bfS_x)^2 < \epsilon$
    } \,, \\
    \infty & \text{otherwise}\,,
  \end{cases}
\end{equation}
with $\beta_\con = N/f_\con$.  As has been shown in \cite{drastic}, 
the constraint $(\p_\mu \bfS_x)^2<\epsilon$ is effectively
replaced by the $\delta$-function $\delta\big( (\p_\mu \bfS_x)^2 -
\epsilon\big)$ in leading order in $1/N$.  
As a consequence, the physics for $\epsilon < f_\con/2$ (to this order)
is equivalent to the topological action
\cite{drastic} where only the constraint is present, i.e.
\begin{equation} \label{Atop} 
  \calA_\mathrm{top}[\bfS] =
  \begin{cases}
    0 & \text{for }
    (\p_\mu \bfS_x)^2 < \epsilon\,, \\
    \infty & \text{otherwise}\,.
  \end{cases}
\end{equation}
The mass gap in this case is also given by \eqref{m_st} with $f_\st$ replaced
by $2\epsilon$.  (For $\epsilon > f_\con/2$ the constraint is irrelevant in
leading order, and the physics is given by the standard action, with
$f_\st=f_\con$.  We assume here that $\epsilon< f_\con/2$.)

\subsection{Mixed action}
The mixed action \cite{drastic} is given by 
\begin{equation} \label{Amix} 
  \calA_\mathrm{mix}[\bfS] = \frac{\beta_\m}{2}
  \sum_{x,\mu} (\p_\mu \bfS_x)^2 +\frac{\gamma_\m}{4} \sum_{x,\mu} [(\p_\mu
  \bfS_x)^2]^2 \,.
\end{equation}

The large $N$ limit is taken as
\begin{equation} \label{bg} 
  \beta_\m = \frac{N}{f_\m} \,,\qquad
  \gamma_\m = \frac{2N}{\kappa_\m^2} \,,
\end{equation}
keeping $f_\m$ and $\kappa_\m$ constant.  

The infinite volume mass gap is determined by the effective coupling
$\fhat_\m$ defined by
\begin{equation} \label{fhat} \frac{1}{\fhat_\m} =
  \frac{1}{2f_\m}+\sqrt{\frac{1}{4f_\m^2}+\frac{1}{\kappa_\m^2}} \,,
\end{equation}
and is given by the same expression as \eqref{m_st} with $f_\st$ replaced by
$\fhat_\m$.

Besides $\fhat_\m$ we introduce
\begin{equation} \label{qdef} 
  r_\m = \frac{\kappa_\m}{f_\m} 
  \qquad \mathrm{and}
  \quad q_\m=\frac12\left( r_\m+\sqrt{r_\m^2+4}\right) \,.
\end{equation}
While the mass gap is determined by $\fhat_\m$ alone, 
the cutoff effects depend on the ratio $r_\m$ as well.

It is easy to see that $r_\m=\infty$ ($q_\m=\infty$) 
corresponds to the standard action.  
The choice $r_\m=0$ ($q_\m=1$) gives the ``purely quartic'' 
action (where $\beta_\m=0$). 
It is less obvious that the $r_\m\to -\infty$ ($q_\m\to 0$) 
limit gives the constrained action. For large negative $r_\m$ 
the action density in \eqref{Amix} has a deep minimum at
\begin{equation}
  (\p_\mu \bfS_x)^2 = -\frac{\beta_\m}{\gamma_\m} = \frac12 \fhat_\m
  \Big( 1+ \order{r_\m^{-2}}\Big) \,,
\end{equation}
which shows that this limit indeed corresponds to the constrained action with
$\epsilon=\fhat_\m/2$.

The technical details (introduction of auxiliary variables,
the gap equation, the calculation of the 2-point and 4-point
functions) can be found in \cite{drastic}, and we shall not repeat
them here.

The cutoff dependence of the step scaling function
$\Sigma(2,u,a/L)$ is described by \eqref{uprime}, \eqref{nu_mix}.
The functions $T_i(u)$ are given by
\begin{equation} \label{Ti_def}
  T_i(u) = \frac{1}{f_0'(u'_\infty)}\left( 
      \Phi_i(u) - \frac14 \Phi_i(u'_\infty) \right) \,,
\end{equation}
\begin{equation} \label{Phii_def}
  \begin{aligned}
    \Phi_0(u) & = f_1(u)+\frac18 u^2 f_0(u) 
    - \frac{2f_2(u)-u^2f_0(u)}{1 - \frac{2}{\pi}+q_\m^2}
    \left(u f_0'(u)+\frac{1}{4\pi} \right) \,, \\
    \Phi_1(u) & = - \frac18 u^2 \,, \\
    \Phi_2(u) & = \frac{1}{1+q_\m^2} u^2 \,. \\
   \end{aligned}
\end{equation}
The functions $f_i(u)$ are defined in \eqref{f0u_def}, 
\eqref{f1u}, \eqref{f2u}.
Note that the coefficients $T_i(u)$ depend on the parameter
$q_\m$ defined in \eqref{qdef}. 
As shown above, the coefficients $t_i(u)$ and $\bar t_i(u)$, 
describing the cutoff dependence for the standard and constrained actions 
can be obtained from $T_i(u)$ by setting 
$q_\m=\infty$ and $q_m=0$, respectively.
For the standard action ($q_\m=\infty$) the $z^2$ term is absent.

\section{Matching to the lattice results}
\label{Matching}
As described in subsection~\ref{strategy}, 
we calculate the 2-point and 4-point
correlation functions on the lattice in infinite volume up to
$\order{a^2}$, and compare these with the results obtained
using Symanzik's effective action \eqref{EffLag}.

\subsection{Matching the infinite volume 2-point function}

In leading order in $1/N$ the 2-point function for the mixed action is the
free lattice propagator
\begin{equation}
  \tilde{\mathcal{G}}^{(2)}(k) = \frac{F}{\hat{k}^2+m_0^2} \,,
\end{equation}
where $m_0$ is expressed by the infinite-volume mass gap as
\begin{equation} \label{m0m}
  m_0 = \frac{2}{a} \sinh\frac{m a}{2}\,,
\end{equation}
and $F$ is related to the effective coupling $\fhat_{\mathrm{m}}$ by
\begin{equation} \label{fhatF}
  \fhat_{\mathrm{m}} = F \, \frac{r_{\mathrm{m}}+\sqrt{r_{\mathrm{m}}^2+
4-4a^2m_0^2/F}}{r_{\mathrm{m}}+\sqrt{r_{\mathrm{m}}^2+4}}
  = F + \order{a^2m_0^2} = F + \order{\exp(-4\pi/F)}\,.
\end{equation}

The gap equation is
\begin{equation} \label{gapeq_m}
  \frac{1}{F} = \int_{-\pi/a}^{\pi/a} \frac{\mrd^2 q}{(2\pi)^2}
  \frac{1}{\hat{q}^2+m_0^2} = z (1+\psi a^2m_0^2) + \order{a^4m_0^4} \,,
\end{equation}
where $z$ is defined in \eqref{zdef} and
\begin{equation} \label{Psi}
  \psi=-\frac18 +\frac{1}{96\pi z}\,.
\end{equation}

We now have
\begin{equation}
  G_\mathrm{latt}^{(2)}(k) = \frac{1}{z}(k^2+m^2) 
  + \frac{a^2}{z}\left\{ \frac{1}{12}(k^4-m^4)-\psi m^2(k^2+m^2)
  \right\} + \order{a^4}\,.
\end{equation}
According to Symanzik's conjecture this should be equal to
\begin{equation}
  y^2 \left\{
    G_R^{(2)}(k) 
    + a^2 \sum_i v_i G_{i(R)}^{(2)}(k) +\order{a^4}
    \right\} \,.
\end{equation}
Using (\ref{G2})
we find that the scaling parts match if the wave function renormalization
constant is of the form
\begin{equation}
  y^2 = \frac{1}{z f} \left( 1 + w_8 a^2m_0^2\right) + \order{a^4m_0^4}\,.
\end{equation}
Matching the $\order{a^2}$ terms gives
\begin{equation}
  \frac{1}{12}(k^4-m^4)-\psi m^2(k^2+m^2)
  = w_8 m^2(k^2+m^2)+ \frac{1}{f}\sum_i v_i G_{i(R)}^{(2)}(k) \,.
\end{equation}
Using the formulas \eqref{G2_1}-\eqref{G2_9} we find
\begin{align}
  v_6 & = \frac{1}{24} \,, \label{v6} \\
  \frac{v_3}{2} + v_5 & = \frac{1}{16} \,, \label{v3v5}\\
  v_5 + \frac{v_8}{2m^2}  & = -\frac{1}{16} - \psi - w_8 \,, \label{v5v8}\\
  v_1 - \frac{2 v_9}{m^2} & = - \frac{f}{96\pi} \,. \label{v1v9}
\end{align}

\subsection{Matching the infinite volume 4-point function (standard action)}

We analyse first the ST case. In this case the 4-point function is
\begin{equation}
  \tilde{\cal G}^{(4)}_\mathrm{latt}=-\frac{2f^2_\st}
  {N B_\mathrm{latt}(p)}
  \frac{1}{\hat p_1^2+m_0^2}
  \frac{1}{\hat p_2^2+m_0^2}
  \frac{1}{\hat q_1^2+m_0^2}
  \frac{1}{\hat q_2^2+m_0^2} \,,
\end{equation}
where $p=p_1+p_2=-q_1-q_2$ and 
\begin{equation} \label{B_latt}
  B_\mathrm{latt}(p)=\int_{-\pi/a}^{\pi/a}\frac{{\rm d}^2 q}{(2\pi)^2}
  \frac{1}{(\hat q^2+m_0^2)\left[m_0^2+(\widehat{p+q})^2\right]} \,.
\end{equation}
Expanding in the lattice spacing we have 
\begin{equation} \label{B_latt_exp}
  B_\mathrm{latt}(p)
  = b(p)+a^2 b_1(p)+\order{a^4} \,.
\end{equation}
The functions $b(p)$ and $b_1(p)$ are given in appendix~C.

The amputated 4-point function on the lattice is then
\begin{equation}
  \begin{split}
    G^{(4)}_{\rm latt}=&-\frac{2}{Nz^2 b(p)}+\frac{2a^2}
    {Nz^2 b(p)}\Bigg\{2\psi m^2+\frac{b_1(p)}{b(p)}\\
    &-\frac{1}{12}\left[
      \frac{p_1^4-m^4}{p_1^2+m^2}
      +\frac{p_2^4-m^4}{p_2^2+m^2}
      +\frac{q_1^4-m^4}{q_1^2+m^2}
      +\frac{q_2^4-m^4}{q_2^2+m^2}\right]
    \Bigg\}+\order{a^4} \,,
  \end{split}
\end{equation}
which we have to match to the DR result,
\begin{equation}
  \frac{1}{z^2f^2}\left(1+2w_8a^2m_0^2+\order{a^4m_0^4}\right)
  \left\{G^{(4)}_R+a^2\sum_iv_iG^{(4)}_{i(R)}+
    \order{a^4} \right\} \,.
\end{equation}

Using (\ref{G4}) we see that
the scaling pieces indeed agree and from the rest we have 
\begin{equation}
  \begin{split}
    2\psi m^2+\frac{ b_1(p)}{b(p)}  
    &-\frac{1}{12}\left[
      \frac{p_1^4-m^4}{p_1^2+m^2}
      +\frac{p_2^4-m^4}{p_2^2+m^2}
      +\frac{q_1^4-m^4}{q_1^2+m^2}
      +\frac{q_2^4-m^4}{q_2^2+m^2}\right]=\\
    &-2w_8m^2+\frac{N b(p)}{2f^2}\sum_iv_iG^{(4)}_{i(R)}.
  \end{split}
\end{equation}
Here we have to use the results \eqref{G4_1}-\eqref{G4_9}.
Because contributions of some operators have momentum dependence 
not occurring in the above expression we first have 
\begin{equation} \label{v2v4v7}
  v_2=v_4=v_7=0.
\end{equation}
Next we get 
\begin{equation}  \label{v3}
  v_3=\frac{1}{4} \,,
\end{equation}
which combined with the results of the previous subsection gives 
\begin{equation}  \label{v5_v8}
  v_5=-\frac{1}{16}\qquad {\rm and}\quad v_8=-2m^2(\psi+w_8) \,.
\end{equation}
The last matching equation is
\begin{equation} \label{last}
\frac{v_1}{2f}=b_1(p)-\frac{m^4}{12}\frac{\partial b(p)}
{\partial m^2}-\frac16 Y \,.
\end{equation}
After inserting the expressions for $b(p)$ and $b_1(p)$
given in appendix~C, we obtain the simple result
\begin{equation} \label{v1}
  v_1 = \frac{f}{4}\left(z-\frac{1}{8\pi}\right)
  =\frac{1}{4}+\frac{f}{16\pi}\left(
    \ln\frac{8}{\pi}-\frac{1}{2}-\Gamma^\prime(1)\right) \,. \\
\end{equation}

Eqs.~\eqref{v6}, \eqref{v2v4v7}, \eqref{v1} are consistent with (3.30) 
from \cite{perturbative} where we found to first order in coupling 
constant PT\footnote{Note the extra $N$ factor in our $U_1$
  with respect to $U_1$ of that paper.}: 
\begin{equation}
  \begin{aligned}
    & v_1=\frac{1}{4}+\order{f} \,, \quad
    v_2= \order{f} \,, \quad
    v_3 = \frac14 + \order{f} \,, \quad
    v_4 = \order{f} \,, \quad \\
    & v_5 = -\frac{1}{16} + \order{f} \,, \quad
    v_6=\frac{1}{24}+\order{f} \,, \quad
    v_7=\order{f} \,.
  \end{aligned}
\end{equation}

\subsection{Matching the infinite volume 4-point function (mixed action)}

Referring to appendix~B of \cite{drastic} the mixed case amounts
to adding an extra piece to the propagator of the auxiliary 
field,
\begin{equation} \label{XXtr}
  \widetilde{\triangle}(p) \to 
  \widetilde{\triangle}(p) + 
  a^2 \sum_{\mu,\nu}X_\mu(p_1,p_2)X_\nu(q_1,q_2)\widetilde{\triangle}_{\mu\nu}(p)
\end{equation}
where 
\begin{align}
  &\widetilde{\triangle}(p) = \frac{1}{B_\mathrm{latt}(p)} \,, \\
  &\widetilde{\triangle}_{\mu\nu}(p)\sim
  \frac{2\pi}{F^2} 
  \left[k_1\left(\delta_{\mu\nu}-\frac12\right)+k_2\right]+\order{a^2}\,,
  \\
  k_1&=\frac{1}{\pi-2+\pi q_{\mathrm{m}}^2}\,,\qquad 
  k_2=\frac{1}{2\pi(1+q_{\mathrm{m}}^2)} \,, \label{k1k2}
\end{align}
and
\begin{equation}
  X_\mu(p_1,p_2) = -\hat{p}_{1\mu}\hat{p}_{2\mu}
  - \frac{h_\mu(p)}{h(p)} \,.
\end{equation}
Here $h(p)= B_{\rm latt}(p)$
and $h_\mu(p)$ are given in \eqref{B_latt}, \eqref{hmup}.

Expressing the terms with specific momentum dependence by the 
appropriate combinations of $G_{i(R)}^{(4)}$, 
after some algebra we get for the additional part of the 4-point
function of the mixed action
\begin{equation}
  \begin{aligned}
    &\sum_{\mu,\nu}X_\mu(p_1,p_2)X_\nu(q_1,q_2)\widetilde{\triangle}_{\mu\nu}(p)
    =
    \frac{2\pi N}{F^2}  \Bigl\{
    \frac{k_1}{8f^3}\left(G_{7(R)}^{(4)}-2G_{2(R)}^{(4)}\right)
    \\ 
    &+k_2\left(
      \frac{z^2}{f}\left[G_{1(R)}^{(4)}+\frac{m^2}{2}G_{9(R)}^{(4)}\right]
      +\frac{z}{f^2}G_{3(R)}^{(4)}+\frac{1}{f^3}G_{4(R)}^{(4)}
      +\frac{z m^2}{f^2}G_{8(R)}^{(4)}\right)\Bigr\}
    + \order{a^2} \,.
  \end{aligned}
\end{equation}
The matching gives
\begin{equation}
  v_i^{\mathrm{mix}}=v_i^{\mathrm{st}}+x_i\,,\,\,\,\,i\ne5\,,
\end{equation}
where the $x_i$ are determined through
\begin{equation}
  \sum_{i\ne5} x_iG_{i(R)}^{(4)}
  =-\frac{2f^2F^2}{N}
  \sum_{\mu,\nu}X_\mu(p_1,p_2)X_\nu(q_1,q_2)\widetilde{\triangle}_{\mu\nu}(p)
  + \order{a^2} \,, 
\end{equation}
and so we get
\begin{align} \label{v1m}
  v_1^{\mathrm{mix}}&=f\left[-4\pi z^2 k_2
    +\frac{z}{4}-\frac{1}{32\pi}\right]\,,
  \\ \label{v2m}
  v_2^{\mathrm{mix}}&=\frac{\pi k_1}{f}\,,
  \\ \label{v3m}
  v_3^{\mathrm{mix}}&=\frac14-4\pi z k_2\,,
  \\ \label{v4m}
  v_4^{\mathrm{mix}}&=-\frac{4\pi k_2}{f}\,,
  \\ \label{v5m}
  v_5^{\mathrm{mix}}&=-\frac{1}{16} + 2\pi z k_2 \,, 
  \\ \label{v6m}
  v_6^{\mathrm{mix}}&=\frac{1}{24}\,,
  \\ \label{v7m}
  v_7^{\mathrm{mix}}&=-\frac{\pi k_1}{2f}\,,
  \\ \label{v8m}
  v_8^{\mathrm{mix}}&=-2m^2\left[\psi+2\pi z k_2+w_8\right]\,,  
  \\ \label{v9m}
  v_9^{\mathrm{mix}}&= \frac{m^2}{2}\left[\frac{f}{96\pi}
    +v_1^{\mathrm{mix}}\right]\,.
\end{align}
The coefficient $v_8^{\mathrm{mix}}$ can be determined only up to the
parameter $w_8$ which can be fixed if we first impose a renormalization
condition on the lattice 2-point functions.

Note that the matching coefficients for the mixed action
are non-perturbative in the coupling constant, some of them
contain a $1/f$ factor. 
This explains why the leading artifact in this case
is $a^2\ln^2 a$ as opposed to $a^2\ln a$ for the standard action.
The mixed action is perturbative (i.e. the quadratic part dominates)
only when $f_\m \ll \kappa_\m$, i.e. $q_{\mathrm{m}}\gg 1$. 
In this case the $q_{\mathrm{m}}^2$ in the denominator of \eqref{k1k2}
compensates the $1/f$ factor in the continuum limit.

To minimize the lattice artifacts for the step scaling function 
one has to take \cite{drastic} $q_{\mathrm{m}}^2=8/\hat{f}_\m + \order{1}$. 
Inserting this and $z=1/f + \order{1}$, and 
$1/\hat{f}_\m=1/f + \order{1}$ into \eqref{v1m}-\eqref{v9m} 
the coefficients $v_1$ and $v_3$ become $\order{f}$ while for 
the standard action they are $\order{1}$.
As has been shown in \cite{perturbative}, indeed the operator
$U_1$ is responsible for the leading lattice artifacts.

\subsection{The matching for the step scaling function}
Let us parametrize the \lq\lq couplings'' ${\bar g}_i$ defined in
\eqref{gbar_1}-\eqref{gbar_9} as
\begin{equation}
  {\bar g}_i=2f^{e_i}M^4(L)\frac{\Psi_i(u)}{u^3f^\prime_0(u)}.
\end{equation}

Using also the finite volume results from appendix~B of 
\cite{drastic}, we have
\begin{align}
  e_1& = 0,\qquad \Psi_1(u)=-\frac{u^2}{2} \,, \\
  e_2& = 2,\qquad \Psi_2(u)=-\frac{1}{2}\left(
    \frac{1}{4\pi}+uf^\prime_0(u)\right)\left(2f_2(u)-u^2f_0(u)\right) \,, \\
  e_6& = 1,\qquad \Psi_6(u)=24f_1(u)+3u^2f_0(u)-\frac{3u^2}{8\pi} \,, \\
  e_7& = 2,\qquad \Psi_7(u)=-2\Psi_2(u) \,, \\
  e_9& = 0,\qquad \Psi_9(u)=L^2 \,,
\end{align}
and 
\begin{equation}
  \Psi_3(u)=\Psi_4(u)=\Psi_5(u)=\Psi_8(u)=0.
\end{equation}
It is interesting to observe that the off-shell operators
$U_3$, $U_4$, $U_5$ and $U_8$ do not contribute to the finite volume mass.
$U_9$ does contribute, but since $\Psi_9(u_\infty^\prime)=(2L)^2$, its
contribution to the step scaling function also vanishes:
\begin{equation}
  \Psi_9(u)-\frac{1}{4}\Psi_9(u_\infty^\prime)=0.
\end{equation}

Using the mass formula \eqref{Mlatt} we find 
\begin{equation}
  \nu(u,z)=\frac{1}{f^\prime_0(u_\infty^\prime)}\left\{
    \Psi(u)-\frac{1}{4}\Psi(u_\infty^\prime)\right\}
\end{equation}
with 
\begin{equation}
  \Psi(u)=\frac{v_1}{f}\Psi_1(u)+f(v_2-2v_7)\Psi_2(u)+v_6\Psi_6(u).
\end{equation}

Inserting the coefficients $v_i$ given by eqs.~\eqref{v1m}-\eqref{v9m}
one obtains for the mixed action\footnote{The results for the standard and 
constrained actions can be obtained from here be setting $q_{\mathrm{m}}^2$ 
to $\infty$ and $0$, respectively.}
\begin{align}
  \label{v1_mix}
  v_1 & = f\left(-\frac{2z^2}{1+q_{\mathrm{m}}^2}
    +\frac{z}{4}-\frac{1}{32\pi}\right) \,, \\
  \label{v2_mix}
  v_2-2v_7 & = \frac{2\pi}{\pi-2+\pi q_{\mathrm{m}}^2}\frac{1}{f} \,, \\
  \label{v6_mix}
  v_6 & = \frac{1}{24} \,.
\end{align}
These results agree with those of \cite{drastic}.
Note that matching the finite volume mass gap does not fix completely
the coefficients $v_i$ -- it yields only the relations given above.
Also note that the coefficients  of the effective action 
obtained by matching in finite volume do not depend on the volume,
as expected by general considerations \cite{GL,HL}.

\section{Conclusions}
\label{conclusions}
The extrapolation of lattice data to the continuum limit
is a crucial step in the determination of low energy 
physical quantities in Quantum Chromodynamics.
Our present theoretical understanding of lattice artifacts
is based on considerations of Symanzik. He argued that
asymptotically the leading ultraviolet cutoff effects are 
described by an effective continuum Lagrangian containing 
a finite number of higher dimensional operators, restricted 
by symmetries only of the underlying lattice regularization,
and with coefficients depending on the lattice spacing.

In this paper we have given additional support for
Symanzik's theory of lattice artifacts, by determining
the local effective Lagrangian for a class of lattice actions
of the 2-dimensional nonlinear O($N$) sigma model in a 
non-perturbative setting of the $1/N$-expansion at the leading order.
The class of models considered includes also ones which
are considered to belong to the same universality class
as standard actions, but for which the usual perturbative expansions 
are (as yet) not available.

The effective actions are shown to reproduce previously computed
lattice artifacts of the associated step scaling functions 
which are defined in finite volume. 
Once established the effective action can be used to predict
cutoff effects for various observables.

\section*{Acknowledgments}

This work is supported in 
part by funds provided by the Schweizerischer Nationalfonds (SNF). 
The ``Albert Einstein Center for Fundamental Physics'' at Bern University 
is supported by the ``Innovations- und Kooperationsprojekt C-13'' 
of the Schweizerische Uni\-ver\-si\-t\"ats\-kon\-fe\-renz (SUK/CRUS).
J.~B. and F.~N. thank the MPI Munich, where part of this work has been
done, for hospitality. 
This investigation has been supported in part by the Hungarian
National Science Fund OTKA (under K 77400).

\begin{appendix}

\section{Two-point functions with operator insertion 
  in finite volume using dimensional regularization}
\label{AppA}
\subsubsection*{Ordinary 2-point function and source-dependent operators}

From the leading order large $N$ Feynman rules we see that the sigma field
propagator is of the free form
\begin{equation}
  \overline{G}^{(2)}=f_0(k^2+M^2).
\end{equation}
We see that the wave function renormalization constant $Z$ is the same as
appearing in the coupling renormalization and the renormalized 2-point 
function is
\begin{equation}
  \overline{G}^{(2)}_{(R)}=f(k^2+M^2).
\end{equation}
The correlators of the source-dependent operators $U_3$, $U_4$ and $U_5$
are given in \cite{perturbative}. The correlators of $U_3$ are simply related
to the ordinary correlation functions (without operator insertion):
\begin{equation}
  \tilde{\cal G}^{(r)}_3(p_1,\dots,p_r)=\frac{1}{4}\left(\sum_{k=1}^r\,p_k^2
  \right)\,\tilde{\cal G}^{(r)}(p_1,\dots,p_r).
\end{equation}
From this representation it is clear that we have for the operator 
renormalization constant $Z_3=1$.

The correlators of $U_4$ are related \cite{perturbative} to those of
the isospin tensor operator $S^aS^b-\frac{1}{N}\delta^{ab}$. Its 2-point
correlation function vanishes and therefore the operator renormalization
constant $Z_4$ can only be calculated from its 4-point function.

The operator $U_5$ corresponds to a contact term: its 2-point function
is given by
\begin{equation}
  \overline{G}^{(2)}_5=f_0(k^2+M^2)^2
\end{equation}
and all higher correlation functions vanish: $\overline{G}^{(r)}_5=0$ for
$r>2$. We see that $Z_5=1$ and the renormalized 2-point function is
\begin{equation}
  \overline{G}^{(2)}_{5(R)}=f(k^2+M^2)^2.
\end{equation}
Finally for the source-dependent operator $U_8$ we have
\begin{equation}
  \tilde{\cal G}^{(r)}_8=\frac{r}{4}\,\tilde{\cal G}^{(r)}
\end{equation}
for all $r$. From the $r=2$ case we obtain
\begin{equation}
  \overline{G}^{(2)}_8=\frac{f_0}{2}(k^2+M^2),\qquad\quad Z_8=1,\qquad\quad
  \overline{G}^{(2)}_{8(R)}=\frac{f}{2}(k^2+M^2).
\end{equation}

\subsubsection*{The coupling-related operator $U_9$}

This operator is related to the integral of the action density 
\begin{equation}
  \frac{1}{2g_0^2}\int{\rm d}^Dx\,\partial_\mu S\cdot\partial_\mu S
\end{equation}
and if we denote the corresponding correlation functions by 
${\cal G}^X_{{\cal A}}$, where $X$ is any combination of operators, we 
can derive the following Ward identity:
\begin{equation}
  {\cal G}^X_{{\cal A}}=g_0^2\frac{\partial}{\partial g_0^2}\,{\cal G}^X=
  f_0\frac{\partial}{\partial f_0}\,{\cal G}^X.
\label{Ward}
\end{equation}
Applying this to the 2-point function we get
\begin{equation}
  \tilde{\cal G}^{(2)}_{{\cal A}}=\frac{f_0}{k^2+M^2}-\frac{1}{\overline{B}(0)
    (k^2+M^2)^2},
\end{equation}
where we used the relation
\begin{equation}
  f_0\frac{\partial M^2}{\partial f_0}=\frac{1}{f_0\overline{B}(0)},
\end{equation}
which can be obtained from the gap equation (\ref{gapM}). Here 
$\overline{B}(0)$ denotes the finite volume version of the integral
(\ref{Bp}) evaluated at $p=0$:
\begin{equation}
  \overline{B}(0)=\int_{(L)}\frac{{\rm d}^D q}{(2\pi)^D}\frac{1}{(q^2+M^2)^2}.
\end{equation}
Using the definition (\ref{U9def}) we find
\begin{equation}
  \overline{G}^{(2)}_9=-\frac{1}{\overline{B}(0)}.
\end{equation}
From this we see that $Z_9=Z$ and
\begin{equation}
  \overline{G}^{(2)}_{9(R)}=-\frac{1}{\overline{b}(0)}.
\end{equation}

\subsubsection*{The operators $U_1$, $U_2$}

Let us introduce the index notation $\dot{1}$, $\dot{2}$ for correlation
functions with insertion of the space integral of the local operators
\begin{equation}
  \frac{1}{g_0^2}\,{\cal O}_i,\qquad\qquad i=1,2.
\label{dot}
\end{equation}
We then have for the 2-point functions (see figures \ref{F1}, \ref{F2}):
\begin{figure}
  \begin{center}
    \psset{unit=0.7mm}
    \begin{pspicture}(0,0)(90,40)
      \qdisk(12,15){2}
      \qdisk(12,19){2}
      \rput[l]{45}(19,12){\psellipse(10,10)(10,4)}
      \rput(12,5){\psellipse(10,10)(10,4)}
      \rput(3,17){$\mathcal{O}_{1,2}$}
      \rput(31,15){%
        \psCoil[coilaspect=0,coilheight=1.5,coilwidth=3]{90}{1890}
      }
      \psline{-}(55,15)(80,25)
      \psline{-}(55,15)(80,5)
      \rput(83,25){$k$}
      \rput(84,5){$-k$}
    \end{pspicture}
  \end{center}
  \begin{caption}   {\label{F1}}
    A contribution of the 2-point function with insertion
    of the operator $\mathcal{O}_1$ or $\mathcal{O}_2$. The two black
    blobs denote the two sigma-field contractions in \eqref{O1_O3},
    and have similar meaning in all other figures.
    The isospin indices of the outgoing legs are contracted.
  \end{caption}
\end{figure}
\begin{figure}
  \begin{center}
    \psset{unit=0.7mm}
    \begin{pspicture}(0,0)(60,40)
      \qdisk(12,15){2}
      \qdisk(12,19){2}
      \rput[l]{45}(19,12){\psellipse(10,10)(10,4)}
      \rput(12,5){\psellipticarc{-}(30,10)(30,6){90}{-90}}
      \rput(3,17){$\mathcal{O}_{1,2}$}
      \rput(47,21){$k$}
      \rput(48,9){$-k$}
    \end{pspicture}
    \begin{caption}   {\label{F2}}
    \end{caption}
  \end{center}
\end{figure}

\begin{eqnarray}
  \overline{G}^{(2)}_{\dot1}&=&\frac{f_0^2}{2}\left(\sum_{\mu=1}^D
    {\cal H}_\mu\right)\,\left(k^2-\frac{1}{\overline{B}(0)}\,\sum_{\nu=1}^D
    {\cal L}_\nu\right),\\
  \overline{G}^{(2)}_{\dot2}&=&\frac{f_0^2}{2}\sum_{\mu=1}^D
  {\cal H}_\mu\,\left(k_\mu^2-\frac{{\cal L}_\mu}{\overline{B}(0)}\right),
\end{eqnarray}
where
\begin{equation}
  {\cal H}_\mu=\int_{(L)}\frac{{\rm d}^D q}{(2\pi)^D}\,\frac{q_\mu^2}
  {q^2+M^2},\qquad\quad
  {\cal L}_\mu=\int_{(L)}\frac{{\rm d}^D q}{(2\pi)^D}\,\frac{q_\mu^2}
  {(q^2+M^2)^2}.
\end{equation}
These integrals can be expressed in terms of the differences
\begin{equation}
  {\cal H}_1-{\cal H}_2=h,\qquad\qquad\qquad
  {\cal L}_1-{\cal L}_2=\ell
\end{equation}
and sums
\begin{equation}
  \sum_{\mu=1}^D{\cal H}_\mu=-\frac{M^2}{f_0},\qquad\qquad
  \sum_{\mu=1}^D{\cal L}_\mu=\frac{1}{f_0}-M^2\overline{B}(0).
\end{equation}
Taking into account that there is just one momentum component ($q_1$)
with discrete spectrum and $D-1$ continuous ones, we have 
\begin{equation} \label{method1}
  \sum_{\mu=1}^D{\cal H}_\mu=D{\cal H}_2+h
\end{equation}
and similarly for ${\cal L}_\mu$.

Using the above relations we can now write
\begin{equation}
  \overline{G}^{(2)}_{\dot1}=-\frac{M^2f_0}{2}\left(k^2-\frac{1}{\overline{B}(0)}
    \sum_{\nu=1}^D{\cal L}_\nu\right)=-\frac{M^2f_0}{2}(k^2+M^2)+\frac{M^2}
  {2\overline{B}(0)}
\end{equation}
leading to (using (\ref{U12def}))
\begin{equation}
  \overline{G}^{(2)}_1=\frac{M^2}{2\overline{B}(0)},\qquad\qquad
  Z_1=Z,\qquad\qquad
  \overline{G}^{(2)}_{1(R)}=\frac{M^2}{2\overline{b}(0)}.
\end{equation}

An alternative way to express $\overline{G}^{(2)}_{\dot1}$ is
\begin{equation}
  \overline{G}^{(2)}_{\dot1}=\frac{f_0^2}{2}\left\{D{\cal H}_2k^2+hk^2-
    \frac{D^2{\cal H}_2{\cal L}_2}{\overline{B}(0)}-\frac
    {D(h{\cal L}_2+\ell{\cal H}_2)}{\overline{B}(0)}-\frac{\ell h}
    {\overline{B}(0)}\right\}.
\end{equation}
Comparing to
\begin{equation}
  \overline{G}^{(2)}_{\dot2}=\frac{f_0^2}{2}\left\{{\cal H}_2k^2+hk_1^2-
    \frac{D{\cal H}_2{\cal L}_2}{\overline{B}(0)}-\frac
    {h{\cal L}_2+\ell{\cal H}_2}{\overline{B}(0)}-\frac{\ell h}
    {\overline{B}(0)}\right\}
\end{equation}
we obtain using the definition in (\ref{U12def})
\begin{equation}
  \overline{G}^{(2)}_2=\frac{f_0^2h}{2}\left\{k^2-Dk_1^2+(D-1)\frac{\ell}
    {\overline{B}(0)}\right\}.
\label{method1res}
\end{equation}
Here we can demonstrate that our results are independent of the way we
regularize the finite volume problem. Indeed, if, instead of (\ref{method1})
we write
\begin{equation} \label{method2}
  \sum_{\mu=1}^D{\cal H}_\mu=D{\cal H}_2+(D-1)h
\end{equation}
corresponding to $D-1$ discrete and one continuous momentum components,
we find instead of (\ref{method1res})
\begin{equation} \label{method2res}
  \overline{G}^{(2)}_2=\frac{f_0^2h}{2}\left\{Dk_2^2-k^2+(D-1)\frac{\ell}
    {\overline{B}(0)}\right\}.
\end{equation}
In the $D\to2$ limit, both (\ref{method1res}) and (\ref{method2res})
lead to
\begin{equation}
  Z_2=\frac{1}{Z},\qquad\qquad
  \overline{G}^{(2)}_{2(R)}=\frac{f^2h}{2}\left\{k_2^2-k_1^2+\frac{\ell}
    {\overline{b}(0)}\right\}.
\end{equation}
The integrals (\ref{hell}) corresponding to the differences $h$ and $\ell$
are finite.

\subsubsection*{The tensor operators $U_6$, $U_7$}

The main part of $U_7$ is the term
\begin{equation}
  \frac{N}{f_0}\,\sum_{\mu=1}^D\left(\partial_\mu S\cdot\partial_\mu S\right)^2
\end{equation}
and corresponds to the 2-point function 
\begin{equation}
  4f_0^2\sum_{\mu=1}^D{\cal H}_\mu\left(k_\mu^2-\frac{{\cal L}_\mu}
    {\overline{B}(0)}\right)=8\overline{G}^{(2)}_{\dot2}.
\end{equation}
Combining this with the remaining contributions from (\ref{U7def})
we find
\begin{equation}
  \overline{G}^{(2)}_7=\left(8-\frac{16}{D+2}\right)\,\overline{G}^{(2)}_{\dot2}
  -\frac{8}{D+2}\,\overline{G}^{(2)}_{\dot1}=-\frac{8}{D+2}\,\overline{G}^{(2)}_2
\end{equation}
and hence
\begin{equation}
  Z_7=\frac{1}{Z},\qquad\qquad \overline{G}^{(2)}_{7(R)}=-
  2\overline{G}^{(2)}_{2(R)}=f^2h\left\{k_1^2-k_2^2-\frac{\ell}{\overline{b}(0)}
  \right\}.
\end{equation}

The main part of $U_6$ is 
\begin{figure}
  \begin{center}
    \psset{unit=0.7mm}
    \begin{pspicture}(0,0)(60,30)
      \qdisk(12,15){2}
      \rput(12,5){\psellipticarc{-}(30,10)(30,6){90}{-90}}
      \rput(5,15){$Q_6$}
      \rput(47,21){$k$}
      \rput(48,9){$-k$}
    \end{pspicture}
    \begin{caption}   {\label{F3}}
    \end{caption}
  \end{center}
\end{figure}
\begin{figure}
  \begin{center}
    \psset{unit=0.7mm}
    \begin{pspicture}(0,0)(90,30)
      \qdisk(12,15){2}
      \rput(12,5){\psellipse(10,10)(10,10)}
      \rput(5,15){$Q_6$}
      \rput(31,15){%
        \psCoil[coilaspect=0,coilheight=1.5,coilwidth=3]{90}{1890}
      }
      \psline{-}(55,15)(80,25)
      \psline{-}(55,15)(80,5)
      \rput(83,25){$k$}
      \rput(84,5){$-k$}
    \end{pspicture}
    \begin{caption}   {\label{F4}}
    \end{caption}
  \end{center}
\end{figure}

\begin{equation} \label{Q6def}
  Q_6 = \frac{N}{f_0}\sum_{\mu=1}^DS\cdot\partial_\mu^4S
\end{equation}
and corresponds to the 2-point function (see figures \ref{F3},\ref{F4})
\begin{equation}
  2f_0\left(\sum_{\mu=1}^D\,k_\mu^4-\frac{1}{\overline{B}(0)}\sum_{\mu=1}^D
    {\cal K}_\mu\right),
\end{equation}
where
\begin{equation}
  {\cal K}_\mu=\int_{(L)}\frac{{\rm d}^D q}{(2\pi)^D}\,\frac{q_\mu^4}
  {(q^2+M^2)^2}. 
\end{equation}
Using the definition (\ref{U6def}) the 2-point function of $U_6$ becomes
\begin{equation}
  \overline{G}^{(2)}_6=2f_0\sum_{\mu=1}^D\,k_\mu^4+\frac{6f_0}{D+2}\,\left(
    M^4-(k^2)^2\right)-\frac{2f_0}{\overline{B}(0)}\,\overline{{\cal R}},
\end{equation}
with
\begin{equation}
  \overline{{\cal R}}=\sum_{\mu=1}^D{\cal K}_\mu+\frac{6M^2}{D+2}\,\frac{1}{f_0}
  =\overline{\rho}+{\rm O}(\varepsilon).
\end{equation}
The final result for this operator is 
\begin{equation}
  Z_6=1,\qquad\qquad
  \overline{G}^{(2)}_{6(R)}=2fk^4+\frac{3f}{2}(M^4-(k^2)^2)
  -\frac{2f\,\overline{\rho}}{\overline{b}(0)}.
\end{equation}

To obtain the infinite volume results \eqref{G2}-\eqref{G2_9} from the above
formulae is straightforward. One has to use the $L\to\infty$ limits
\begin{equation}
  b(0)=\frac{1}{4\pi m^2}\qquad\qquad{\rm and}\qquad\qquad
  \rho=\frac{3m^2}{16\pi}.
\end{equation}

\subsubsection*{Calculation of the finite volume integrals}

By introducing a Feynman parameter, we can rewrite the gap equation
\begin{equation}
  \frac{1}{f_0}=\int_0^\infty{\rm d}t\,{\rm e}^{-tM^2}\,\int_{(L)}
  \frac{{\rm d}^D q}{(2\pi)^D}\,{\rm e}^{-tq^2}
\end{equation}
and similarly the basic loop integral at zero momentum:
\begin{equation}
  \overline{B}(0)=\int_0^\infty t\,{\rm d}t\,{\rm e}^{-tM^2}\,\int_{(L)}
  \frac{{\rm d}^D q}{(2\pi)^D}\,{\rm e}^{-tq^2}.
\end{equation}
Let us concentrate on the contribution of a single one dimensional integral 
in the last $D$-dimensional integral:
\begin{equation}
  \int_{-\infty}^\infty\,\frac{{\rm d}q}{2\pi}\,{\rm e}^{-tq^2}=\frac{1}
  {\sqrt{4\pi t}}.
\end{equation}
The corresponding \lq\lq integral'', if the momentum variable is discrete,
is really an infinite sum of the form
\begin{equation}
  \frac{1}{L}\sum_{n=-\infty}^\infty\,{\rm e}^{-t\left(\frac{2\pi n}{L}
    \right)^2}.
\label{discrete}
\end{equation}
Here and in all other similar sums we encounter
the theta-function $S(x)$ defined by
\begin{equation}
  S(x)=\sum_{n=-\infty}^\infty\,{\rm e}^{-\pi xn^2}.
\end{equation}
Using this notation, (\ref{discrete}) can be written as 
\begin{equation}
  \frac{1}{L}\,S\left(\frac{4\pi t}{L^2}\right).
\end{equation}
A useful identity satisfied by the function $S(x)$ is
\begin{equation}
  S(x)=\frac{1}{\sqrt{x}}\,S\left(\frac{1}{x}\right),
\end{equation}
which can be proven using the Poisson resummation formula. Thus an alternative
form of (\ref{discrete}) is
\begin{equation}
  \frac{1}{\sqrt{4\pi t}}\,S\left(\frac{L^2}{4\pi t}\right).
\end{equation}

Now we rewrite the gap equation and the zero momentum loop integral as
\begin{align}
  \frac{1}{f_0}&=\int_0^\infty\,{\rm d}t\,{\rm e}^{-tM^2}\,
  \frac{1}{(4\pi t)^{D/2}}
  \,S\left(\frac{L^2}{4\pi t}\right)
  =\int_0^\infty\,{\rm d}t\,{\rm e}^{-tm^2}\,\frac{1}{(4\pi t)^{D/2}},
  \label{gapuj}\\
  \overline{B}(0)&=\int_0^\infty\,t\,{\rm d}t\,{\rm e}^{-tM^2}\,
  \frac{1}{(4\pi t)^{D/2}}\,S\left(\frac{L^2}{4\pi t}\right).\label{B0uj}
\end{align}
The advantage of using the infinite sum $S(x)$ with this argument
is that the $n=0$ term is identical with the infinite volume limit
of the given quantity, while the $n\not=0$ terms are UV finite \cite{HL}.

From (\ref{gapuj}) we obtain the equation
\begin{equation}
  \int_0^\infty{\rm d}t\,\left({\rm e}^{-tM^2}
    -{\rm e}^{-tm^2}\right)\,\frac{1}{(4\pi t)^{D/2}}
  +\int_0^\infty{\rm d}t\,{\rm e}^{-tM^2}\frac{1}{(4\pi t)^{D/2}}
  \left[S\left(\frac{L^2}{4\pi t}\right)-1\right]=0.
\end{equation}
Both integrals have finite $D\to2$ limits and evaluating the first one
and using the definition (\ref{Fdef}) in the second one leads to the final 
result (\ref{fvm}) for calculating the LWW coupling $u=LM$.
The expression (\ref{B0uj}) is also finite for $D\to2$ and gives
\begin{equation}
  4\pi M^2\,\overline{b}(0)=u^2\int_0^\infty{\rm d}t\,{\rm e}^{-tu^2}
  \left\{1+2\sum_{k=1}^\infty\,{\rm e}^{-\frac{k^2}{4t}}\right\}=1-uF^\prime (u).
\end{equation}

Next we consider integrals of the form
\begin{equation}
  \int_{(L)}\frac{{\rm d}^D q}{(2\pi)^D}\,\frac{q^2_\mu}{(q^2+M^2)^{\sigma+1}}
  =\frac{1}{\Gamma(\sigma+1)}\,
  \int_0^\infty t^\sigma{\rm d}t\,{\rm e}^{-tM^2}\int_{(L)}
  \frac{{\rm d}^D q}{(2\pi)^D}\,q_\mu^2\,{\rm e}^{-tq^2}.
\end{equation}
We will need integrals with $\sigma=0,1$. If $\mu\not=1$ then the last integral
is simply
\begin{equation}
  \frac{1}{2t}\,\frac{1}{(4\pi t)^{D/2}}\,S\left(\frac{L^2}{4\pi t}\right)
\end{equation}
whereas for $\mu=1$ it is
\begin{equation} 
  \frac{1}{2t}\,\frac{1}{(4\pi t)^{D/2}}\,S\left(\frac{L^2}{4\pi t}\right)
  -\frac{1}{(4\pi t)^{D/2}}\,\frac{\partial}{\partial t}
  S\left(\frac{L^2}{4\pi t}\right).
\end{equation}
To evaluate $h$ and $\ell$ defined in (\ref{hell}) we only need to integrate
the difference between the above two terms, which is finite as $D\to2$:
\begin{equation}
\begin{split}
  -\int_0^\infty t^\sigma{\rm d}t\,&{\rm e}^{-tM^2}\,\frac{1}{4\pi t}\,
  \frac{\partial}{\partial t}\left[S\left(\frac{L^2}{4\pi t}\right)-1\right]\\
  &=-\frac{(L^2)^{\sigma-1}}{2\pi}\int_0^\infty{\rm d}t\,{\rm e}^{-tu^2}
  \left\{t^{\sigma-1}u^2+(1-\sigma)t^{\sigma-2}\right\}\sum_{k=1}^\infty
  \,{\rm e}^{-\frac{k^2}{4t}}.
\end{split}
\end{equation}
For $\sigma=1$ we obtain from this the first
and for $\sigma=0$ the second relation in (\ref{landh}).

Finally we consider
\begin{equation}
  \int_{(L)}\frac{{\rm d}^D q}{(2\pi)^D}\,\frac{q^4_\mu}{(q^2+M^2)^2}
  =\int_0^\infty t\,{\rm d}t\,{\rm e}^{-tM^2}\int_{(L)}
  \frac{{\rm d}^D q}{(2\pi)^D}\,q_\mu^4\,{\rm e}^{-tq^2}.
\end{equation}
Here for $\mu\not=1$ the last integral is
\begin{equation}
  \frac{3}{4t^2}\,\frac{1}{(4\pi t)^{D/2}}\,S\left(\frac{L^2}{4\pi t}\right)
\end{equation}
whereas for $\mu=1$ it is
\begin{equation}
  \frac{1}{(4\pi t)^{D/2}}\left\{\frac{3}{4t^2}\,S\left(\frac{L^2}{4\pi t}\right)
    -\frac{1}{t}\frac{\partial}{\partial t} S\left(\frac{L^2}{4\pi t}\right)+
    \frac{\partial^2}{\partial t^2} S\left(\frac{L^2}{4\pi t}\right)\right\}.
\end{equation}
Summing over $\mu$, as required by the first term in (\ref{barrho}),
we get
\begin{equation}
  \int_0^\infty t\,{\rm d}t{\rm e}^{-tM^2}\,
  \frac{1}{(4\pi t)^{D/2}}\left\{\frac{3D}{4t^2}\,
    S\left(\frac{L^2}{4\pi t}\right)
    -\frac{1}{t}\frac{\partial}{\partial t} S\left(\frac{L^2}{4\pi t}\right)+
    \frac{\partial^2}{\partial t^2} S\left(\frac{L^2}{4\pi t}\right)\right\}.
\end{equation}
This has to be combined with the second term in (\ref{barrho}), which
is of the form
\begin{equation}
  \frac{6M^2}{D+2}\,
  \int_0^\infty {\rm d}t{\rm e}^{-tM^2}\,
  \frac{1}{(4\pi t)^{D/2}}\,S\left(\frac{L^2}{4\pi t}\right).
\end{equation}
Let us now separate the contributions into a \lq\lq finite'' part, which is
obtained by the replacement
\begin{equation}
  S\left(\frac{L^2}{4\pi t}\right)\longrightarrow
  S\left(\frac{L^2}{4\pi t}\right)-1
\end{equation}
corresponding to UV finite integrals and an \lq\lq infinite'' part,
which corresponds to the substitution
\begin{equation}
  S\left(\frac{L^2}{4\pi t}\right)\longrightarrow1.
\end{equation}
We first evaluate the \lq\lq infinite'' contributions. We find
\begin{equation}
\begin{split}
  &\frac{3D}{4(4\pi)^{D/2}}\int_0^\infty{\rm d}t\,{\rm e}^{-tM^2}\,t^{-1-{D/2}}
  +\frac{6M^2}{(D+2)(4\pi)^{D/2}}
  \int_0^\infty{\rm d}t\,{\rm e}^{-tM^2}\,t^{-{D/2}}\\
  &=\frac{3M^D}{(4\pi)^{D/2}}
  \left\{\frac{D}{4}\Gamma\left(-{D/2}\right)+
    \frac{2}{D+2}\Gamma\left(1-{D/2}\right)\right\}
  =\frac{3M^D}{(D+2)(4\pi)^{D/2}}
  \Gamma\left(2-{D/2}\right).
\end{split}
\end{equation}
In the $D\to2$ limit this is finite and gives $3M^2/16\pi$.
The \lq\lq finite'' part gives 
\begin{equation}
  \begin{split}
    \int_0^\infty t\,{\rm d}t&\,{\rm e}^{-tM^2}\,
    \frac{1}{4\pi t}\left\{\frac{3}{2t^2}\,
      \left[S\left(\frac{L^2}{4\pi t}\right)-1\right]
      -\frac{1}{t}\frac{\partial}{\partial t} S\left(\frac{L^2}{4\pi t}\right)+
      \frac{\partial^2}{\partial t^2} S\left(\frac{L^2}{4\pi t}\right)\right\}\\
    &+\frac{3M^2}{2}\,
    \int_0^\infty {\rm d}t{\rm e}^{-tM^2}\,
    \frac{1}{4\pi t}\,\left[S\left(\frac{L^2}{4\pi t}\right)-1\right]\\
    &=\frac{M^2}{2\pi}\int_0^\infty{\rm d}t\,{\rm e}^{-tu^2}
    \left(\sum_{k=1}^\infty{\rm e}^{-\frac{k^2}{4t}}\right)
    \left\{\frac{1}{2u^2t^2}+\frac{1}{2t}
      +u^2\right\}\\
    &=-\frac{M^2}{2\pi u^2}H(u)+\frac{M^2}{4\pi}F(u)-\frac{M^2}{4\pi}uF^\prime(u)=
    -M^2\ell-\frac{h}{2}.
  \end{split}
\end{equation}
Adding the two contributions we finally have the result (\ref{barrhores}).

\section{Four-point functions with operator insertion 
in infinite volume using dimensional regularization}
\label{AppB}
\subsubsection*{Ordinary 4-point function and source-dependent operators}

The leading large $N$ 4-point functions are of order $1/N$. Using the Feynman
rules given in section~\ref{SymTheo}, 
the leading order ordinary 4-point function is
\begin{equation}
  G^{(4)}=-\frac{2f_0^2}{NB(p)}.
\end{equation}
After renormalization we get for the renormalized 4-point function
(\ref{G4}).
The 4-point function of the source-dependent operator $U_3$ can be obtained
from the above by multiplying it by a momentum-dependent factor: 
\begin{equation}
  G^{(4)}_3=-\frac{f_0^2}{2NB(p)}(p_1^2+p_2^2+q_1^2+q_2^2)
\end{equation}
and after renormalization this becomes (\ref{G4_3}).
Correlation functions of the source-dependent operator $U_4$ can be calculated
using the rules given in \cite{perturbative}. For the 4-point function we get
\begin{equation}
  G^{(4)}_4=\frac{f_0^3}{N}\left(\frac{p_1^2+p_2^2}{2}+m^2\right)
  \left(\frac{q_1^2+q_2^2}{2}+m^2\right).
\end{equation}
This allows us to determine $Z_4$ (the only operator renormalization
constant not determined by the 2-point functions). We obtain $Z_4=1/Z$
and (\ref{G4_4}).
Next we recall that 
\begin{equation}
  G^{(4)}_5=G^{(4)}_{5(R)}=0,
\end{equation}
since all $r>2$ correlation functions of $U_5$ vanish. Finally the 4-point
correlators of $U_8$ coincide with the ordinary ones:
\begin{equation}
  G^{(4)}_8=-\frac{2f_0^2}{NB(p)},\qquad\qquad\quad
  G^{(4)}_{8(R)}=-\frac{2f^2}{Nb(p)}.
\end{equation}

\subsubsection*{The action-related operator $U_9$}

From the Ward identity (\ref{Ward}) we obtain
\begin{equation}
  G^{(4)}_{{\cal A}}=-\frac{4f_0^2}{NB(p)}+\frac{2f_0}{N}X^{(o)},
\end{equation}
where the coefficient $X^{(o)}$ is the bare version of (\ref{renX}):
\begin{equation}
  X^{(o)}=\frac{1}{B(p)B(0)}\left\{\frac{1}{B(p)}\frac{\partial B(p)}
    {\partial m^2}+
    \frac{1}{p_1^2+m^2}+\frac{1}{p_2^2+m^2}+\frac{1}{q_1^2+m^2}+
    \frac{1}{q_2^2+m^2}\right\}=X+\order{\varepsilon}.
\end{equation}
Subtracting the $U_8$ part we get
\begin{equation}
  G^{(4)}_9=\frac{2f_0}{N}X^{(o)}
\end{equation}
and after renormalization (\ref{G4_9}).

\subsubsection*{The operators $U_1$, $U_2$}

There are two types of contributions to the 4-point functions 
$G^{(4)}_{\dot1}$ and $G^{(4)}_{\dot2}$ (recall these are
correlators of the operators (\ref{dot})): either the four external legs are
attached to the four legs of the local operators, or all four external
legs are coupled to one pair of operator legs and the other pair of operator 
legs are contracted with each other. The latter type of contributions are 
proportional to the action-related 4-point function $G^{(4)}_{{\cal A}}$. 
Indicating the first type of contributions by a hat, we have
\begin{equation}
  G^{(4)}_{\dot1}=\hat G^{(4)}_{\dot1}-\frac{m^2}{2}G^{(4)}_{{\cal A}},
  \qquad\qquad
  G^{(4)}_{\dot2}=\hat G^{(4)}_{\dot2}-\frac{m^2}{2D}G^{(4)}_{{\cal A}}.
\end{equation}
The \lq\lq hatted'' contributions are constructed from the building blocks
shown in figures \ref{F5}-\ref{F7}
\begin{figure}
  \begin{center}
    \psset{unit=0.7mm}
    \begin{pspicture}(0,0)(50,40)
      \qdisk(12,15){2}
      \qdisk(12,19){2}
      \rput(12,10){\psellipticarc{-}(30,10)(30,10){90}{180}}
      \rput(12,6){\psellipticarc{-}(30,10)(30,10){180}{270}}
      \psline{-}(12,15)(42,15)
      \psline{-}(12,19)(42,19)
      \rput(46,14){$q_1$}
      \rput(46,5){$q_2$}
      \rput(46,29){$p_1$}
      \rput(46,20){$p_2$}
      \rput(3,17){$\mathcal{O}_{1,2}$}
    \end{pspicture}
    \begin{caption}   {\label{F5}}
    \end{caption}
  \end{center}
\end{figure}

\begin{figure}
  \begin{center}
    \psset{unit=0.7mm}
    \begin{pspicture}(0,0)(80,45)
      \qdisk(12,15){2}
      \qdisk(12,19){2}
      \rput[l]{45}(19,12){\psellipse(10,10)(10,4)}
      \rput(24,35){%
        \psCoil[coilaspect=0,coilheight=1.5,coilwidth=3]{180}{1890}
      }
      \psline{-}(48,35)(68,40)
      \psline{-}(48,35)(68,30)
      \rput(12,6){\psellipticarc{-}(56,10)(56,10){180}{270}}
      \psline{-}(12,15)(68,15)
      \rput(72,15){$q_1$}
      \rput(72,5){$q_2$}
      \rput(72,40){$p_1$}
      \rput(72,30){$p_2$}
      \rput(38,40){$p$}
      \rput(3,17){$\mathcal{O}_{1,2}$}
    \end{pspicture}
    \begin{caption}   {\label{F6}}
    \end{caption}
  \end{center}
\end{figure}

\begin{figure}
  \begin{center}
    \psset{unit=0.7mm}
    \begin{pspicture}(0,-5)(80,40)
      \qdisk(12,15){2}
      \qdisk(12,19){2}
      \rput[l]{30}(16,10){\psellipse(10,10)(10,4)}
      \rput[l]{-30}(7,5){\psellipse(10,10)(10,4)}
      
      \rput(3,17){$\mathcal{O}_{1,2}$}
      \rput(28,28){%
        \psCoil[coilaspect=0,coilheight=1.5,coilwidth=3]{90}{1890}
      }
      \psline{-}(52,28)(72,34)
      \psline{-}(52,28)(72,22)
      
      \rput(28,5){%
        \psCoil[coilaspect=0,coilheight=1.5,coilwidth=3]{90}{1890}
      }
      \psline{-}(52,5)(72,11)
      \psline{-}(52,5)(72,-1)
      
      \rput(76,34){$p_1$}
      \rput(76,22){$p_2$}
      \rput(76,11){$q_1$}
      \rput(76,-1){$q_2$}
      \rput(40,33){$p$}
      \rput(40,10){$-p$}
    \end{pspicture}
    \begin{caption}   {\label{F7}}
    \end{caption}
  \end{center}
\end{figure}

\begin{align}
  \hat G^{(4)}_{\dot1}&=\frac{f_0^3}{4N}\,
  T_{\mu\mu}(p_1,p_2)\,T_{\nu\nu}(q_1,q_2),\label{Gdot1}\\
  \hat G^{(4)}_{\dot2}&=\frac{f_0^3}{4N}\,
  T_{\mu\nu}(p_1,p_2)\,T_{\mu\nu}(q_1,q_2),\label{Gdot2}
\end{align}
where 
\begin{equation}
  -T_{\mu\nu}(p_1,p_2)=p_{1\mu}p_{2\nu}+p_{1\nu}p_{2\mu}+\frac{2}{B(p)}\left(
    A_0\,\delta_{\mu\nu}+B_0\,p_\mu p_\nu\right).
\end{equation}
The $p$-dependent scalars $A_0$, $B_0$ are defined by the regularized loop 
integral
\begin{equation}
  \int\frac{{\rm d}^Dq}{(2\pi)^D}\,\frac{q_\mu(q+p)_\nu}{(q^2+m^2)
    [m^2+(q+p)^2]}=A_0\,\delta_{\mu\nu}+B_0\,p_\mu p_\nu
\end{equation}
and are explicitly given as
\begin{equation}
  \begin{split}
    A_0 & = \frac{1}{2(D-1)f_0}-\frac{m^2+p^2/4}{D-1}B(p),\\
    p^2\,B_0 & = \frac{D-2}{2(D-1)f_0}+\left[\frac{m^2}{D-1}
      +\frac{p^2(2-D)}{4(D-1)} \right]B(p).
  \end{split}
\end{equation}
$A_0$ is divergent but $B_0$ has a finite $D\to2$ limit:
\begin{equation}
  B_0=b_0+\order{\varepsilon},\qquad\qquad
  p^2b_0=-\frac{1}{4\pi}+m^2b(p).
\end{equation}
Using
\begin{equation}
  DA_0+p^2\,B_0=\frac{1}{f_0}-(m^2+p^2/2)\,B(p) \,,
\end{equation}
$\hat G^{(4)}_{\dot1}$ can be simplified:
\begin{equation}
  \begin{split}
    \hat G^{(4)}_{\dot1} & =\frac{f_0^3}{N}
    \left(m^2+\frac{p_1^2+p_2^2}{2}-\frac{1}{f_0B(p)}\right)
    \left(m^2+\frac{q_1^2+q_2^2}{2}-\frac{1}{f_0B(p)}\right)\\
    & = \frac{f_0}{NB^2(p)}-\frac{2m^2f_0^2}{NB(p)}+G^{(4)}_3+G^{(4)}_4.
  \end{split}
\end{equation}
Comparing the last line and the definition in (\ref{U12def})
we see that the 4-point function of $U_1$ simplifies to
\begin{equation}
  G^{(4)}_1=\frac{f_0}{NB^2(p)}-\frac{m^2f_0}{N}X^{(o)}.
\end{equation}
The renormalized version is obtained by the substitution $f_0\to f$ and
replacing the rest by its finite $D\to2$ limit as given in (\ref{G4_1}).

There is an alternative representation of $\hat G^{(4)}_{\dot1}$ which will
be useful later. Going back to the definition (\ref{Gdot1}) we can write
\begin{equation}
  \hat G^{(4)}_{\dot1}=\frac{f_0^3}{4N}\left\{
    \frac{4D^2A_0^2}{B^2(p)}+\frac{4DA_0}{B(p)}\left[p_1p_2+q_1q_2+\frac{2B_0p^2}
      {B(p)}\right]+{\cal F}_{\dot1}\right\}
\end{equation}
where
\begin{equation}
  {\cal F}_{\dot1}=
  \frac{4B_0^2(p^2)^2}{B^2(p)}+\frac{4B_0p^2}{B(p)}(p_1p_2+q_1q_2)
  +4(p_1p_2)(q_1q_2).
\end{equation}
Similarly, from (\ref{Gdot2}) we calculate
\begin{equation}
  \hat G^{(4)}_{\dot2}=\frac{f_0^3}{4N}\left\{
    \frac{4DA_0^2}{B^2(p)}+\frac{4A_0}{B(p)}\left[p_1p_2+q_1q_2+\frac{2B_0p^2}
      {B(p)}\right]+{\cal F}_{\dot2}\right\}
\end{equation}
with
\begin{equation}
  {\cal F}_{\dot2}=
  \frac{4B_0^2(p^2)^2}{B^2(p)}+\frac{4B_0}{B(p)}\left[(p_1p)(p_2p)+
    (q_1p)(q_2p)\right]+2(p_1q_1)(p_2q_2)+2(p_1q_2)(p_2q_1).
\end{equation}
To compute the 4-point function of $U_2$ we first observe that the
action-related terms cancel and we can start from
\begin{equation}
  G^{(4)}_2=\hat G^{(4)}_{\dot1}-D\hat G^{(4)}_{\dot2}.
\end{equation}
After some algebra we see that the divergent $A_0$-dependent terms also 
cancel and we are left with
\begin{equation}
  \begin{split}
    G^{(4)}_2 & = \frac{f_0^3}{N}\Bigg\{-(D-1)
    \left(\frac{B_0p^2}{B(p)}\right)^2
    +\frac{B_0p^2}{B(p)}\Big[p_1p_2+q_1q_2-\frac{D}{p^2}(p_1p)(p_2p)\\
    &-\frac{D}{p^2}(q_1p)(q_2p)\Big]
    +(p_1p_2)(q_1q_2)-\frac{D}{2}(p_1q_1)(p_2q_2)
    -\frac{D}{2}(p_1q_2)(p_2q_1) \Bigg\}.
  \end{split}
\end{equation}
The renormalized 4-point function (\ref{G4_2}) is obtained from this
in the $D\to2$ limit by making the substitutions $f_0\to f$ and
\begin{equation}
  \frac{B_0p^2}{B(p)}\longrightarrow\frac{b_0p^2}{b(p)}
  =m^2-\frac{1}{4\pi b(p)} =Q.
\end{equation}

\subsubsection*{The tensor operator $U_7$}

The essential part of $U_7$ is the operator
\begin{equation}
  \frac{N}{f_0}\sum_{\mu=1}^D(\partial_\mu S\cdot \partial_\mu S)^2
\end{equation}
and we will denote its 4-point function by $G^{(4)}_{\dot7}$. Very similarly
to the previously discussed cases, we can separate this 
4-point function into two parts:
\begin{equation}
  G^{(4)}_{\dot7}=\hat G^{(4)}_{\dot7}-\frac{4m^2}{D}G^{(4)}_{{\cal A}},
\end{equation}
where
\begin{equation}
  \hat G^{(4)}_{\dot7}=\frac{8f_0^3}{N}\sum_{\mu=1}^D\left(
    p_{1\mu}p_{2\mu}+\frac{A_0}{B(p)}+\frac{B_0p_\mu^2}{B(p)}\right)
  \left(q_{1\mu}q_{2\mu}+\frac{A_0}{B(p)}+\frac{B_0p_\mu^2}{B(p)}\right).
\end{equation}
This can be simplified to
\begin{equation}
  \hat G^{(4)}_{\dot7}=\frac{8f_0^3}{N}\left\{
    \frac{DA_0^2}{B^2(p)}+\frac{A_0}{B(p)}\left[p_1p_2+q_1q_2+\frac{2B_0p^2}
      {B(p)}\right]+{\cal F}_{\dot7}\right\}
\end{equation}
where
\begin{equation}
  {\cal F}_{\dot7}=\sum_{\mu=1}^D
  \left(p_{1\mu}p_{2\mu}+\frac{B_0p_\mu^2}{B(p)}\right)
  \left(q_{1\mu}q_{2\mu}+\frac{B_0p_\mu^2}{B(p)}\right).
\end{equation}
Using (\ref{U7def}) we notice that action-dependent terms again cancel
and we can write
\begin{equation}
  G^{(4)}_7=\hat G^{(4)}_{\dot7}-\frac{8}{D+2}\hat G^{(4)}_{\dot1}
  -\frac{16}{D+2}\hat G^{(4)}_{\dot2}.
\end{equation}
The divergent, $A_0$-dependent terms also cancel here and we are left with
\begin{equation}
  G^{(4)}_7=\frac{f_0^3}{N}\left(8{\cal F}_{\dot7}-\frac{2}{D+2}{\cal F}_{\dot1}
    -\frac{4}{D+2}{\cal F}_{\dot2}\right).
\end{equation}
The final result (\ref{G4_7}) is obtained from this last formula by making
the replacement $f_0\to f$ and writing the $D\to2$ limit of the rest as
a quadratic polynomial in $Q$.

\subsubsection*{The tensor operator $U_6$}

Here we will use the following representation 
\cite{perturbative}\footnote{This follows from 
formulae (A.10) and (A.34) of that paper in the large $N$ limit.}
of the operator $U_6$:
\begin{equation}
  U_6=Q_6-\frac{6}{D+2}\left(U_5-4U_4+\frac{4N}{f_0}{\cal O}_1\right) \,.
\end{equation}
Here $Q_6$, defined in \eqref{Q6def},
is the main part whose correlation functions we will denote by the dotted
index $\dot6$. Using the fact that $U_5$ is a pure contact term from the
above representation we have
\begin{equation}
  G^{(4)}_6=G^{(4)}_{\dot6}-\frac{24}{D+2}\left(G^{(4)}_{\dot1}-G^{(4)}_4\right).
\end{equation}

There are several contributions to the four-point correlation function 
$G^{(4)}_{\dot6}$. 
The corresponding Feynman diagrams are shown in 
figures~\ref{F8}-\ref{F11}.

\begin{figure}
  \begin{center}
    \psset{unit=0.7mm}
    \begin{pspicture}(0,0)(80,46)
      \qdisk(12,15){2}
      \rput(12,5){\psellipticarc{-}(60,10)(60,6){90}{-90}}
      \rput(5,15){$Q_6$}
      \rput{45}(35,19){%
        \psCoil[coilaspect=0,coilheight=1.5,coilwidth=3]%
        {180}{1890}
      }
      \psline{-}(51.5,36)(72,42)
      \psline{-}(51.5,36)(72,30)
      \rput(76,42){$p_1$}
      \rput(76,30){$p_2$}
      \rput(76,21){$q_1$}
      \rput(76,9){$q_2$}
      \rput(41,31){$p$}
    \end{pspicture}
    \begin{caption}   {\label{F8}}
    \end{caption}
  \end{center}
\end{figure}

\begin{figure}
  \begin{center}
    \psset{unit=0.6mm}
    \begin{pspicture}(0,0)(86,60)
      \qdisk(12,30){2}
      \rput(5,30){$Q_6$}
      \rput(12,30){\psellipse(16,0)(16,16)}
      \rput{30}(40,40){%
        \psCoil[coilaspect=0,coilheight=1.5,coilwidth=3]{0}{1870}
      }
      \psline{-}(60,52)(80,58)
      \psline{-}(60,52)(80,46)
      \rput{-30}(40,20){%
        \psCoil[coilaspect=0,coilheight=1.5,coilwidth=3]{10}{1800}
      }
      \psline{-}(60,10)(80,16)
      \psline{-}(60,10)(80,4)
      
      \rput(84,58){$p_1$}
      \rput(84,46){$p_2$}
      \rput(84,16){$q_1$}
      \rput(84,4){$q_2$}
      \rput(46,50){$p$}
      \rput(46,10){$-p$}
    \end{pspicture}
    \begin{caption}  {\label{F9}}
    \end{caption}
  \end{center}
\end{figure}

\begin{figure}
  \begin{center}
    \psset{unit=0.6mm}
    \begin{pspicture}(0,14)(103,60)
      \qdisk(12,30){2}
      \rput(5,30){$Q_6$}
      \rput(12,30){\psellipse(12,0)(12,12)}
      \rput(35,30){%
        \psCoil[coilaspect=0,coilheight=1.5,coilwidth=3]{70}{1910}
      }
      \rput(59,30){\psellipticarc{-}(40,0)(40,8){90}{-90}}
      \rput{60}(70,34.5){%
        \psCoil[coilaspect=0,coilheight=1.5,coilwidth=3]{180}{1520}
      }
      \psline{-}(79,51)(99,57)
      \psline{-}(79,51)(99,45)
      \rput(103,57){$p_1$}
      \rput(103,45){$p_2$}
      \rput(103,38){$q_1$}
      \rput(103,22){$q_2$}
      \rput(72,45){$p$}
    \end{pspicture}
    \begin{caption} {\label{F10}}
    \end{caption}
  \end{center}
\end{figure}

\begin{figure}
  \begin{center}
    \psset{unit=0.6mm}
    \begin{pspicture}(0,0)(120,60)
      \qdisk(12,30){2}
      \rput(5,30){$Q_6$}
      \rput(12,30){\psellipse(12,0)(12,12)}
      \rput(35,30){%
        \psCoil[coilaspect=0,coilheight=1.5,coilwidth=3]{70}{1900}
      }
      \psline{-}(59,30)(72,42)
      \psline{-}(59,30)(72,18)
      \psline{-}(72,18)(72,42)
      \rput(68.5,18){%
        \psCoil[coilaspect=0,coilheight=1.5,coilwidth=3]{270}{1900}
      }
      \psline{-}(92,18)(110,26)
      \psline{-}(92,18)(110,10)
      \rput(68.5,42){%
        \psCoil[coilaspect=0,coilheight=1.5,coilwidth=3]{270}{1900}
      }
      \psline{-}(92,42)(110,50)
      \psline{-}(92,42)(110,34)
      \rput(114,50){$p_1$}
      \rput(114,34){$p_2$}
      \rput(114,26){$q_1$}
      \rput(114,10){$q_2$}
      \rput(82,48){$p$}
      \rput(82,13){$-p$}
    \end{pspicture}
    \begin{caption} {\label{F11}}
    \end{caption}
  \end{center}
\end{figure}

The sum of the contributions of the Feynman diagrams shown in 
figure~\ref{F8} is
\begin{equation}
  -\frac{4f_0^2}{NB(p)}\sum_{\mu=1}^D\left\{
    \frac{p_{1\mu}^4}{p_1^2+m^2}+\frac{p_{2\mu}^4}{p_2^2+m^2}+
    \frac{q_{1\mu}^4}{q_1^2+m^2}+\frac{q_{2\mu}^4}{q_2^2+m^2}\right\}.
\end{equation}
The \ref{F9} type contribution is
\begin{equation}
  \frac{8f_0^2}{NB^2(p)}\int\,\frac{{\rm d}^Dq}{(2\pi)^D}\,\sum_{\mu=1}^D
  \frac{q_\mu^4}{(q^2+m^2)^2\left[m^2+(q+p)^2\right]}.
\end{equation}
The sum of the \ref{F10} type contributions is
\begin{equation}
  \frac{4f_0^2}{NB(p)B(0)}\left\{
    \frac{1}{p_1^2+m^2}+\frac{1}{p_2^2+m^2}+\frac{1}{q_1^2+m^2}
    +\frac{1}{q_2^2+m^2}\right\}\,
  \sum_{\mu=1}^D\int\,\frac{{\rm d}^Dq}{(2\pi)^D}\,\frac{q_\mu^4}{(q^2+m^2)^2}
\end{equation}
and finally the \ref{F11} type contribution is
\begin{equation}
  \frac{4f_0^2}{NB^2(p)B(0)}\,\frac{\partial B(p)}{\partial m^2}
  \left(\sum_{\mu=1}^D \int\,\frac{{\rm d}^Dq}{(2\pi)^D}\,
    \frac{q_\mu^4}{(q^2+m^2)^2}\right).
\end{equation}
Putting together all contributions we have
\begin{equation}
  \begin{split}
    G^{(4)}_{\dot6} & =-\frac{4f_0^2}{NB(p)}\sum_{\mu=1}^D\left\{
      \frac{p_{1\mu}^4}{p_1^2+m^2}+\frac{p_{2\mu}^4}{p_2^2+m^2}+
      \frac{q_{1\mu}^4}{q_1^2+m^2}+\frac{q_{2\mu}^4}{q_2^2+m^2}\right\}\\
    & +\frac{4f_0^2}{N}X^{(o)} \sum_{\mu=1}^D\int\,
    \frac{{\rm d}^Dq}{(2\pi)^D}\,\frac{q_\mu^4}{(q^2+m^2)^2}\\
    &+\frac{8f_0^2}{NB^2(p)}
    \sum_{\mu=1}^D\int\,\frac{{\rm d}^Dq}{(2\pi)^D}\,\frac{q_\mu^4}
    {(q^2+m^2)^2\left[m^2+(q+p)^2\right]}.
  \end{split}
\end{equation}
Using our earlier results we have
\begin{equation}
  G^{(4)}_{\dot1}-G^{(4)}_4=\frac{f_0}{NB^2(p)}+G^{(4)}_3-\frac{m^2f_0}{N}X^{(o)}
\end{equation}
and finally
\begin{equation}
  \begin{split}
    G^{(4)}_6&=-\frac{4f_0^2}{NB(p)}\sum_{\mu=1}^D\left\{
      \frac{p_{1\mu}^4}{p_1^2+m^2}+\frac{p_{2\mu}^4}{p_2^2+m^2}+
      \frac{q_{1\mu}^4}{q_1^2+m^2}+\frac{q_{2\mu}^4}{q_2^2+m^2}\right\}\\
    &+\frac{12f_0^2}{(D+2)NB(p)}(p_1^2+p_2^2+q_1^2+q_2^2)\\
    &+\frac{4f_0^2}{N}X^{(o)}\left\{
      \sum_{\mu=1}^D\int\,\frac{{\rm d}^Dq}{(2\pi)^D}\,\frac{q_\mu^4}{(q^2+m^2)^2}
      +\frac{6}{D+2}\,\frac{m^2}{f_0}\right\}\\
    &+\frac{8f_0^2}{NB^2(p)}\left\{
      \sum_{\mu=1}^D\int\,\frac{{\rm d}^Dq}{(2\pi)^D}\,\frac{q_\mu^4}
      {(q^2+m^2)^2\left[m^2+(q+p)^2\right]}-\frac{3}{(D+2)f_0}\right\}.
  \end{split}
\end{equation}

The coefficient of $X^{(o)}$ can be simplified if we evaluate
\begin{equation}
  \int\,\frac{{\rm d}^Dq}{(2\pi)^D}\,\frac{q_\alpha q_\beta q_\gamma q_\delta}
  {(q^2+m^2)^2}=\Psi(
  \delta_{\alpha\beta}\delta_{\gamma\delta}+
  \delta_{\alpha\gamma}\delta_{\beta\delta}+
  \delta_{\alpha\delta}\delta_{\beta\gamma}).
\end{equation}
$\Psi$ can be calculated by considering the contraction
\begin{equation}
  D(D+2)\Psi=
  \int\,\frac{{\rm d}^Dq}{(2\pi)^D}\,\frac{(q^2)^2}{(q^2+m^2)^2}=
  -\frac{2m^2}{f_0}+m^4B(0).
\end{equation}
This gives
\begin{equation}
  \int\,\frac{{\rm d}^Dq}{(2\pi)^D}\,\sum_{\mu=1}^D\frac{q_\mu^4}{(q^2+m^2)^2}=
  3D\Psi=
  -\frac{6m^2}{(D+2)f_0}+\frac{3m^4B(0)}{D+2}.
\end{equation}
The four-point function is now simplified to
\begin{equation}
  \begin{split}
    G^{(4)}_6&=-\frac{4f_0^2}{NB(p)}\sum_{\mu=1}^D\left\{
      \frac{p_{1\mu}^4}{p_1^2+m^2}+\frac{p_{2\mu}^4}{p_2^2+m^2}+
      \frac{q_{1\mu}^4}{q_1^2+m^2}+\frac{q_{2\mu}^4}{q_2^2+m^2}\right\}\\
    &+\frac{12f_0^2}{(D+2)NB(p)}(p_1^2+p_2^2+q_1^2+q_2^2)
    +\frac{12m^4f_0^2B(0)}{(D+2)N}X^{(o)}\\
    &+\frac{8f_0^2}{NB^2(p)}Y^{(o)},
  \end{split}
  \label{bareG6}
\end{equation}
where we introduced a second bare coefficient
\begin{equation}
  Y^{(o)}=\sum_{\mu=1}^D\int\,\frac{{\rm d}^Dq}{(2\pi)^D}\,\frac{q_\mu^4}
  {(q^2+m^2)^2\left[m^2+(q+p)^2\right]}-\frac{3}{(D+2)f_0}=
  Y+\order{\varepsilon}.
\end{equation}
The formula (\ref{bareG6}) is now ready for renormalization and for the
renormalized four-point function we get (\ref{G4_6}) of section~\ref{SymTheo}.

It remains to calculate the finite part $Y$. We proceed similarly to the
evaluation of the coefficient of $X^{(o)}$ above but here the calculation is 
more difficult because of the momentum dependence. We start by defining
\begin{equation}
  \begin{split}
    \int\,&\frac{{\rm d}^Dq}{(2\pi)^D}\,\frac{q_\alpha q_\beta q_\gamma q_\delta}
    {(q^2+m^2)^2[m^2+(q+p)^2]}=\Omega_1(
    \delta_{\alpha\beta}\delta_{\gamma\delta}+
    \delta_{\alpha\gamma}\delta_{\beta\delta}+
    \delta_{\alpha\delta}\delta_{\beta\gamma})\\
    &+\Omega_2(
    \delta_{\alpha\beta}p_\gamma p_\delta+
    \delta_{\alpha\gamma}p_\beta p_\delta+
    \delta_{\alpha\delta}p_\beta p_\gamma+
    p_\alpha p_\beta\delta_{\gamma\delta}+
    p_\alpha p_\gamma\delta_{\beta\delta}+
    p_\alpha p_\delta\delta_{\beta\gamma})\\
    &+\Omega_3p_\alpha p_\beta p_\gamma p_\delta.
  \end{split}
  \label{Omegaint}
\end{equation}
What we need is
\begin{equation}
  \sum_{\mu=1}^D\,
  \int\,\frac{{\rm d}^Dq}{(2\pi)^D}\,\frac{q_\mu^4}
  {(q^2+m^2)^2[m^2+(q+p)^2]}=3D\Omega_1+6\Omega_2p^2+\Omega_3\sum_{\mu=1}^D
  p_\mu^4
\end{equation}
and we expect that
\begin{equation}
  \overline{\Omega}=3D\Omega_1+6\Omega_2p^2-\frac{3}{(D+2)f_0}=
  \overline\omega+\order{\varepsilon},\qquad\quad
  \Omega_3=\omega_3+\order{\varepsilon}.
\end{equation}
Anticipating the finiteness of $\overline\omega$ and $\omega_3$ we can write
\begin{equation}
  Y=\overline\omega+\omega_3 p^4.
\end{equation}

For this calculation we have to introduce, in addition to (\ref{Omegaint}),
the following hierarchy of tensor integrals ($r=0,1$):
\begin{equation}
  \begin{split}
    \int\,\frac{{\rm d}^Dq}{(2\pi)^D}\,\frac{q_\alpha q_\beta q_\gamma}
    {(q^2+m^2)^{(1+r)}[m^2+(q+p)^2]}&=K_r(
    \delta_{\alpha\beta}p_\gamma +
    \delta_{\alpha\gamma}p_\beta +
    \delta_{\beta\gamma} p_\alpha)\\
    &+N_r p_\alpha p_\beta p_\gamma,\\
    \int\,\frac{{\rm d}^Dq}{(2\pi)^D}\,\frac{q_\alpha q_\beta}
    {(q^2+m^2)^{(1+r)}[m^2+(q+p)^2]}&=H_r \delta_{\alpha\beta} +
    M_r p_\alpha p_\beta, \\
    \int\,\frac{{\rm d}^Dq}{(2\pi)^D}\,\frac{q_\alpha}
    {(q^2+m^2)^{(1+r)}[m^2+(q+p)^2]}&=\xi_r p_\alpha. \\
  \end{split}
  \label{tensorints}
\end{equation}
By making contraction of the indices with Kronecker deltas and the
momentum vector $p_\mu$ we find the following simple relations among the
tensor integrals:
\begin{equation}
  \begin{aligned}
    &(D+2)\Omega_1+p^2\Omega_2=H_0-m^2 H_1,
    &&(D+4)\Omega_2+p^2\Omega_3=M_0-m^2 M_1,\\
    &2\Omega_1+2p^2\Omega_2=-K_0-p^2K_1,
    &&6\Omega_2+2p^2\Omega_3=-N_0-p^2N_1,\\
    &(D+2)K_0+p^2N_0 =-1/f_0-m^2\xi_0,
    &&(D+2)K_1+p^2N_1=\xi_0-m^2\xi_1,\\
    &2p^2K_0=-p^2H_0, 
    &&4K_0+2p^2N_0=-1/f_0-p^2M_0,\\
    &2p^2K_1=(1/f_0-m^2B(0))/D-H_0-p^2H_1, 
    &&4K_1+2p^2N_1=-M_0-p^2M_1,\\
    &DH_0+p^2M_0=1/f_0-m^2B(p), 
    &&DH_1+p^2M_1=B(p)+(m^2/2)(\partial B(p)/\partial m^2),\\
    &2H_0+2p^2M_0=1/f_0-p^2\xi_0, 
    &&2H_1+2p^2M_1=-\xi_0-p^2\xi_1,\\
    &2p^2\xi_0=-p^2B(p), 
    &&2p^2\xi_1=B(0)-B(p)+(p^2/2)(\partial B(p)/\partial m^2).
  \end{aligned}
\end{equation}
We have solved this overdetermined system of linear relations using
Mathematica. The results for $\Omega_i$ are too bulky to be reproduced here.
In the limit $D\to2$ the formulas simplify enormously and we find 
\begin{eqnarray}
  \overline\omega&=&-3m^2b(p)+\frac{3p^2}{4}\left(m^2+\frac{p^2}{4}\right)\,
  \frac{\partial b(p)}{\partial m^2}+\frac{3}{32\pi}\left(5+\frac{p^2}{m^2}
  \right),\\
  (p^2)^2\omega_3&=&2m^2b(p)
  -\left(\frac{m^4}{2}+m^2p^2+\frac{(p^2)^2}{4}\right)\,
  \frac{\partial b(p)}{\partial m^2}-\frac{1}{8\pi}\left(5+\frac{p^2}{m^2}
  \right).
\end{eqnarray}

\section{Lattice integrals}
\label{AppC}
First note that the continuum limit of the lattice integral 
$B_\mathrm{latt}(p)$ in eq.~\eqref{B_latt} is given by  
$b(p)$ defined in eq.~\eqref{bp} which can be expressed 
analytically as
\begin{equation} \label{bfn}
    b(p) = b(p,m)= \frac{1}{2\pi\sqrt{p^2(p^2+4m^2)}}
    \ln\left(\frac{\sqrt{p^2+4m^2}+\sqrt{p^2}}{\sqrt{p^2+4m^2}
        -\sqrt{p^2}}\right)\,,
\end{equation}
where the explicit reference to the second argument $m$ will 
be used later in this appendix.
$b(p)$ is analytic in $p^2$ with a cut from $-\infty$
to $-4m^2$. Also note that $b(p)\ne0$ for all $p^2$ and
the properties:
\begin{align} \label{bfn_as}
  b(p)&\sim \frac{\ln (p^2/m^2)}{2\pi p^2}\,\,\,\,\,{\rm for}
  \,\,\,p^2\to\infty\,,
  \\ \label{b0}
  b(0)&=\frac{1}{4\pi m^2}\,.
\end{align}
Further we have
\begin{align} \label{dbt_dm}
  \frac{\partial}{\partial m^2}b(p)
  &=-\frac{2}{p^2+4m^2}\left[b(p)+\frac{1}{4\pi m^2}\right]\,,
  \\ \label{dbt_dp}
  \frac{\partial}{\partial p_\mu}b(p)
  &=-\frac{2p_\mu}{p^2(p^2+4m^2)}\left[(p^2+2m^2)b(p)
    -\frac{1}{2\pi}\right]\,.
\end{align}

In order to work out the leading lattice artifacts of
$B_{\rm latt}(p)$ it is convenient to
consider the equivalent representation:
\footnote{before performing the $q_\mu$ integrations we shift the 
  variables
  $q_\mu\to q_\mu+\alpha_\mu$ with $\sin(a\alpha_\mu)
  =-t_2\sin (ap_\mu)T_\mu^{-1}\,,
  \cos(a\alpha_\mu)=\left[t_1+t_2\cos (ap_\mu)\right]T_\mu^{-1}$\,,
  $T_\mu=\sqrt{(t_1+t_2)^2-t_1t_2a^2\hat{p}_\mu^2}$\,.}
\begin{equation}\label{B_lattx}
  \begin{aligned}
    B_{\rm latt}(p)
    &=a^2\int_0^\infty\rmd t_1\rmd t_2\,\rme^{-(t_1+t_2)(a^2m_0^2+4)}
    \prod_{\mu=1}^2 I_0\left(2\sqrt{(t_1+t_2)^2-t_1t_2a^2\hat{p}_\mu^2}\right)
    \\
    &=a^2\int_0^\infty\rmd t\int_0^1\rmd x\,t\rme^{-t(a^2m_0^2+4)}
    \prod_{\mu=1}^2 I_0\left(2tS_\mu(x,p)\right)\,,
  \end{aligned}
\end{equation}
where 
\begin{equation}
  S_\mu(x,p)\equiv\sqrt{1-x(1-x)a^2\hat{p}_\mu^2}\,,
\end{equation}
and $I_r$ denote the modified Bessel functions. 
For $p=0$ we have simply:
\begin{equation}
  B_{\rm latt}(0)=a^2
  \int_0^\infty\rmd t\,t\rme^{-a^2m_0^2t}\left[\rme^{-2t}I_0(2t)\right]^2\,.
\label{B_latt0}
\end{equation}
Noting for large $t$:
\begin{equation}
  \rme^{-2t}I_r(2t) = \frac{1}{\sqrt{4\pi t}}
  \left[1+\frac{1-4r^2}{16t}+\order{t^{-2}}\right]
\end{equation}
and splitting the range of the $t$-integration in (\ref{B_latt0})
into parts $[0,1]$ and $[1,\infty]$ we obtain:
\begin{equation}
  B_{\rm latt}(0)=\frac{1}{4\pi m_0^2}
  +a^2\left[c_1-\frac{1}{32\pi}\ln(a^2m_0^2)\right]
  +\order{a^4m_0^2\ln(a^2m_0^2)}\,, 
\label{Blatt_0}
\end{equation}
with
\begin{align}
  &c_1=-\frac{1}{4\pi}+\sum_{k=1}^4c_1^{(k)}\,,
  \\
  &c_1^{(1)}=\int_0^1\rmd t\,t\left[\rme^{-2t}I_0(2t)\right]^2\,,
  \\
  &c_1^{(2)}=
  \int_1^\infty\rmd t\,\left\{
    t\left[\rme^{-2t}I_0(2t)\right]^2
    -\frac{1}{4\pi}\left(1+\frac{1}{8t}\right)\right\}\,,
  \\
  &c_1^{(3)}=-\frac{1}{32\pi}\int_0^1\rmd t\,t^{-1}(1-\rme^{-t})\,,
  \\
  &c_1^{(4)}=\frac{1}{32\pi}\int_1^\infty\rmd t\,t^{-1}\rme^{-t}\,.
\end{align}
$c_1$ can in fact be determined analytically as 
\begin{equation}
  c_1=\frac{1}{16\pi}\left[\frac52\ln(2)-1\right]\,,
\end{equation}
which can easily be checked numerically.

Proceeding as for the case $p=0$ above:
\begin{align}
  &B_{\rm latt}(p)=
  a^2c_1^{(1)}+a^2\int_1^\infty\rmd t\int_0^1\rmd x\,
  t\rme^{-t(a^2m_0^2+4)}
  \prod_{\mu=1}^2 I_0(2t S_\mu(x,p))
  +\order{a^4}
  \\ 
  &= a^2\left[c_1^{(1)}+c_1^{(2)}\right]
  \nonumber\\
  &+a^2\frac{1}{4\pi}\int_1^\infty\rmd t\int_0^1\rmd x\,
  \frac{\rme^{-t\left[a^2m_0^2+4-2\sum_{\nu=1}^2S_\nu(x,p)\right]}}
       {\prod_{\mu=1}^2S_\mu(x,p)^{1/2}}\left[1+\frac{1}{8t}\right]
  +\order{a^4}
  \\
  &=a^2\left[c_1^{(1)}+c_1^{(2)}\right]
  +\frac{a^2}{4\pi}\int_1^\infty\rmd t\int_0^1\rmd x\,
  \rme^{-ta^2\left[m_0^2+x(1-x)\hat{p}^2\right]}
  \nonumber\\
  & \qquad \times\left\{1+\frac{1}{8t}+\frac14 x(1-x)a^2\hat{p}^2
    -\frac{t}{4}x^2(1-x)^2a^4\hat{p}^4\right\} +\order{a^4}
  \\
   &= b(\hat{p},m_0)+a^2\left[c_1+\sum_{k=1}^3\beta_k\right]
   +\order{a^4}\,,
\end{align}
with
\begin{align}
  \beta_1&=\frac{1}{16\pi}\int_0^1\rmd x\,
  \frac{x(1-x)p^2}{\left[m^2+x(1-x)p^2\right]}
  \label{beta1_def}
  \\
  &=\frac{1}{16\pi}-\frac14 m^2 b(p)\,,
  \\
  \beta_2&=-\frac{1}{16\pi}\int_0^1\rmd x\,
  \frac{x^2(1-x)^2p^4}{\left[m^2+x(1-x)p^2\right]^2}
  \label{beta2_def}
  \\
  &=-\frac{p^4}{4(p^2)^2}
  \left[\frac{1}{4\pi}-2m^2 b(p)
    -m^4\frac{\partial}{\partial m^2} b(p)\right]\,,
  \\
  \beta_3&=-\frac{1}{32\pi}\int_0^1\rmd x\,
  \ln\left[a^2m^2+x(1-x)a^2p^2\right]
  \label{beta3_def}
  \\
  &=-\frac{1}{16}\left[(p^2+4m^2) b(p)
    +\frac{1}{2\pi}\ln(a^2m^2)-\frac{1}{\pi}\right]\,.
\end{align}

Putting the above results together and expanding $m_0$ 
and $\hat{p}$ in $m$ and $p$ respectively (i.e. in $a$) we obtain
\eqref{B_latt_exp} with
\begin{equation}
  b_1(p)=
  \frac{m^4}{12}\frac{\partial}{\partial m^2}b(p)
  -\sum_\mu\frac{p_\mu^3}{24}\frac{\partial}{\partial p_\mu}b(p)
  +c_1+\beta_1+\beta_2+\beta_3 \,.
\end{equation}
Using eqs.~\eqref{dbt_dm},\eqref{dbt_dp} and inserting the resulting 
$b_1(p)$ into eq.~\eqref{last} one indeed obtains
eq.~\eqref{v1}.

For the tadpole (which we need below) one gets 
\begin{align}
  J(m_0)&\equiv\int_{-\pi/a}^{\pi/a}\frac{\rmd^2 q}{(2\pi)^2}
  \frac{1}{m_0^2+\hat{q}^2}
  \\
  &= -\frac{1}{4\pi}\ln(a^2m_0^2)+c_2
  +a^2m_0^2\left[\frac{1}{32\pi}\ln(a^2m_0^2)-\frac{1}{32\pi}-c_1\right]
  +\order{a^4}\,,
  \label{tadpole_exp}
\end{align}
where
\begin{equation}
  c_2=
  \frac{5}{4\pi}\ln(2)=8c_1+\frac{1}{2\pi}\,.
\end{equation}

Next we consider the integrals occurring in the 4-point function
of the mixed action. First
\begin{equation} \label{hmup} 
  \begin{aligned} 
    h_\mu(p)&=\int_{-\pi/a}^{\pi/a}\frac{\rmd^2 q}{(2\pi)^2}
    \frac{\hat{q}_\mu\widehat{(p+q)}_\mu}{(m_0^2+\hat{q}^2)
      \left[m_0^2+\widehat{(p+q)}^2\right]}
    \\
    &=2\cos\left(\frac{ap_\mu}{2}\right)
    \int_0^\infty\rmd t\int_0^1\rmd x\,t\rme^{-t(a^2m_0^2+4)}
    I_0\left(2tS_{\overline{\mu}}(x,p)\right)\times
    \\
    &\times \left[I_0\left(2tS_\mu(x,p)\right)
      -\frac{I_1\left(2tS_\mu(x,p)\right)}{S_\mu(x,p)}\right]\,,
    \\
  \end{aligned}
\end{equation}
where $\overline{1}=2\,,\overline{2}=1\,$.
The expansion for $h_\mu(0)$ to order $a^2$ is immediately obtained
from eq.~(\ref{tadpole_exp}):
\begin{equation}
  \begin{aligned}
    h_\mu(0)&=\frac12\left[J(m_0)
      +m_0^2\frac{\partial}{\partial m_0^2}J(m_0)\right]
    \\
    &= -\frac{1}{8\pi}\ln(a^2m_0^2)+\frac{c_2}{2}-\frac{1}{8\pi}
    +a^2m_0^2\left[\frac{1}{32\pi}\ln(a^2m_0^2)-\frac{1}{64\pi}-c_1\right]
    +\order{a^4}\,.
  \end{aligned}
\end{equation}
Actually for our purposes we only need $h_\mu(p)$ to leading order
\begin{align}
  h_\mu(p)&= 2h^{(1)}_\mu(p)
  -\hat{p}_\mu^2 h^{(2)}_\mu(p)+\order{a^2}\,,
  \\
  h^{(1)}_\mu(p)&=\int_0^\infty\rmd t\int_0^1\rmd x\,t\rme^{-t(a^2m_0^2+4)}
  I_0\left(2tS_{\overline{\mu}}(x,p)\right)
  \nonumber\\
  &\qquad \times \left[I_0\left(2tS_\mu(x,p)\right)
    -I_1\left(2tS_\mu(x,p)\right)\right]\,,
  \\
  h^{(2)}_\mu(p)&=a^2\int_0^\infty\rmd t\int_0^1\rmd x\,t 
   x(1-x)\rme^{-t(a^2m_0^2+4)}
  I_0\left(2tS_{\overline{\mu}}(x,p)\right)I_1\left(2tS_\mu(x,p)\right)\,.
  \nonumber\\
  &
\end{align}
From these representations we can deduce the leading terms
\begin{align}
  h^{(1)}_\mu(p)&= \frac{c_2}{4}-\frac{1}{16\pi}+2\beta_3
  +\order{a^2}\,,  \\
  h^{(2)}_\mu(p)&= \frac{4}{p^2}\beta_1+\order{a^2}\,,
\end{align}
where the $\beta_k$ are defined in 
eqs.~(\ref{beta1_def})-(\ref{beta3_def}). 
Adding the terms we get
\begin{equation}
  h_\mu(p)= \frac{z}{2}+\frac{1}{8\pi}
  -\frac14(p^2+4m^2)b(p)+\frac{p_\mu^2}{p^2}b(p)Q
  +\order{a^2}\,,
\end{equation}
with $Q$, $z$ defined in eqs.~(\ref{Qdef}),(\ref{zdef}) respectively.

Finally for
\begin{equation}
  h_{\mu\nu}(p)=\int_{-\pi/a}^{\pi/a}\frac{\rmd^2 q}{(2\pi)^2}
  \frac{\hat{q}_\mu\widehat{(p+q)}_\mu\hat{q}_\nu\widehat{(p+q)}_\nu}
  {(m_0^2+\hat{q}^2)\left[m_0^2+\widehat{(p+q)}^2\right]}
\end{equation}
appearing in the definition of $\widetilde{\triangle}_{\mu\nu}(p)$
(cf. \cite{drastic}, app.~B) and which we need here only for
$p=0$, $m_0=0$, 
we have
\begin{equation}
  h_{\mu\nu}(0)|_{m_0=0}=\int_{-\pi/a}^{\pi/a}\frac{\rmd^2 q}{(2\pi)^2}
  \frac{\hat{q}_\mu^2\hat{q}_\nu^2}{\left(\hat{q}^2\right)^2}
  =\frac{1}{2\pi a^2}\left[\left(1-\delta_{\mu\nu}\right)
    +(\pi-1)\delta_{\mu\nu}\right]\,.
\end{equation}

\end{appendix}

\end{document}